\documentclass[aps,twocolumn,preprintnumbers,nofootinbib,superscriptaddress]{revtex4-1}
\usepackage{amsmath} \usepackage{graphicx} \usepackage{amsfonts}
\usepackage{array} \usepackage{amsthm} \usepackage{bm}
\usepackage{palatino} \usepackage{mathpazo} 
\usepackage{supertabular}

\usepackage[breaklinks]{hyperref}
\usepackage{color}

\newcommand{\nn}{\nonumber}
\newcommand{\be}{\begin{equation}}
\newcommand{\ee}{\end{equation}}
\newcommand{\ba}{\begin{eqnarray}}
\newcommand{\ea}{\end{eqnarray}}
\newcommand{\bal}{\begin{align}}
\newcommand{\eal}{\end{align}}

\newcommand{\dd}{{\rm d}}

\newcommand{\bb}{\bibitem}

\newcommand{\ka}{\kappa}

\newcommand{\ro}{\rho}

\newcommand{\ta}{\theta}

\newcommand{\Si}{\Sigma}
\newcommand{\De}{\Delta}

\newcommand{\vp}{\varphi}

\newcommand{\bw}{\begin{widetext}}
\newcommand{\ew}{\end{widetext}}

\begin{document}
\title{Epicyclic oscillations of charged particles in stationary solutions immersed in a magnetic field with application to the Kerr--Newman black hole}

\author{Mustapha Azreg-A\"{\i}nou}
\affiliation{Ba\c{s}kent University, Engineering Faculty, Ba\u{g}l\i ca Campus, Ankara, Turkey}


\begin{abstract}
We consider a stationary metric immersed in a uniform magnetic field and determine general expressions for the epicyclic frequencies of charged particles. Applications to the Kerr--Newman black hole is reach of physical consequences and reveals some new effects among which the existence of radially and vertically stable circular orbits in the region enclosed by the event horizon and the so-called {\textquotedblleft innermost\textquotedblright} stable circular orbit in the plane of symmetry.
\end{abstract}


\maketitle

\section{Why studying and determining the epicyclic  oscillations (ECOs)?\label{sec1}}

Any small deviations from a \emph{stable}, geodesic or nongeodesic, motion lead to an epicyclic motion around the stable path. If the epicyclic motion is that of a charged particle, the corresponding frequencies have direct observational effects~\cite{qpos1}-\cite{qpos2}.

The general non-circular motion of charged particles in the geometry of a magnetized Schwarzschild black hole has been investigated~\cite{Frolov} without, however, dealing with the epicyclic oscillations (ECOs). The easiest motion where ECOs may be determined analytically is the circular one~\cite{Tursunov}, from this point of view ECOs of charged particles around circular paths of the Kerr black hole have been studied and their frequencies were determined in terms of the parameters of the black hole, the weak magnetic field in which it is immersed, the charged particle physical properties, and the four-velocity vector of the circular geodesic motion~\cite{Kerr1,Kerr2}. One of the purposes of the present work is to determine the frequencies of the ECOs of charged particles around circular paths of a \emph{general} stationary solution immersed in a magnetic field then apply them to the case of the Kerr--Newman black hole immersed in a magnetic field.

In Sec.~\ref{sec2} we set the ansatz for the metric and electromagnetic field  of a stationary metric immersed in a uniform magnetic field. In Sec.~\ref{pcm} we derive general expressions for the epicyclic frequencies and in Sec.~\ref{ppcm} we discuss some properties of the perturbed circular motion. In Sec.~\ref{secapp} we apply our results to the case of the Kerr--Newman black hole immersed in a magnetic field. We conclude in Sec.~\ref{seccon}.

\section{The metric and the electromagnetic field\label{sec2}}
We split this section into two subsections dealing with the rotating \emph{charged} metric, with charge $Q$, and the linear form of the resulting electromagnetic field due to the interaction of the charge $Q$ with a uniform magnetic field $B$.

\subsection{The metric}
Using the required symmetry properties of a stationary and axisymmetric spacetime that is circular~\cite{Straumann} -- a spacetime admitting the existence of two \emph{commuting}~\footnote{In a circular spacetime, there exists a family of two surfaces everywhere orthogonal
	to the plane defined by the two commuting Killing vectors $\xi^{\mu}_{t}$ and $\xi^{\mu}_{\vp}$. In such a spacetime one may choose the coordinates such that the only nonzero cross term of the metric is $\dd t\dd\vp$.\label{foot1}} Killing vectors, a timelike one $\xi^{\mu}_{t}=(1,0,0,0)$ and a spacelike one $\xi^{\mu}_{\vp}=(0,0,0,1)$, the standard metric of such a spacetime may be brought to the following form in quasi-isotropic coordinates~\cite{rot1,rot2,rot3}:
\begin{equation}\label{g1}
    \dd s^2=-N\dd t^2+A(\dd r^2+r^2\dd\theta^2)+Dr^2\sin^2\theta(\dd\varphi-\Omega\dd t)^2.
\end{equation}
In quasi-isotropic coordinates the equality $g_{rr}=g_{\theta\theta}/r^2$ is justified by the fact that all two-dimensional metrics are related by a conformal factor. Here ($N,\,A,\,D,\,\Omega$) are functions depending on ($r,\theta$) and on the rotation parameter $a$ in a way such that $\Omega\to 0$ as $a\to 0$.

If the rotating solution has one asymptotic region (not a wormhole) and if the \emph{dragging effects are symmetric with respect to the $\theta=\pi/2$ plane} (perpendicular to the axis of rotation), the functions ($N,\,A,\,D,\,\Omega$) assume the following properties by symmetry
\begin{equation}\label{g2}
\partial_{\theta}(N,\,A,\,D,\,\Omega)\big|_{\theta=\pi/2}=(0,\,0,\,0,\,0),
\end{equation}
that is, they are extrema on the $\theta=\pi/2$ plane in a direction perpendicular to it. Not all rotating solutions, with one asymptotic region, are endowed with such a property as are the Kerr--Taub--NUT and Kerr--Newman--Taub--NUT rotating solutions\cite{newman1,newman2}. Due to the gravitomagnetic monopole moment (the NUT charge), these rotating solutions do not drag test particles and fields symmetrically with respect to the $\theta=\pi/2$ plane. There are, however, some wormhole solutions, with two non-symmetrical asymptotic sheets~\cite{worm1,worm2,worm3}, that do well observe the properties~\eqref{g2} but they drag object non-symmetrically with respect to the $\theta=\pi/2$ plane.

From now on we consider only rotating solutions where the properties~\eqref{g2} hold whether these solutions describe black holes, stars, or wormholes. In this case, the only \emph{nonvanishing} elements of $\Gamma^{\mu}_{\alpha\beta}$, $\partial_r\Gamma^{\mu}_{\alpha\beta}$, and $\partial_{\theta}\Gamma^{\mu}_{\alpha\beta}$ \emph{on the $\theta=\pi/2$ plane} are
\begin{align}
\label{Ga}&\Gamma^{t}_{tr},\;\Gamma^{t}_{r\varphi},\;\Gamma^{r}_{tt},\;\Gamma^{r}_{t\varphi},\;\Gamma^{r}_{rr},
\;\Gamma^{r}_{\theta\theta},\;\Gamma^{r}_{\varphi\varphi},\;\Gamma^{\theta}_{r\theta},\;\Gamma^{\varphi}_{tr},\;
\Gamma^{\varphi}_{r\varphi},\\
\label{Gb}&\partial_r\Gamma^{t}_{tr},\;\partial_r\Gamma^{t}_{r\varphi},\;\partial_r\Gamma^{r}_{tt},\;\partial_r\Gamma^{r}_{t\varphi},\;\partial_r\Gamma^{r}_{rr},
\;\partial_r\Gamma^{r}_{\theta\theta},\;\partial_r\Gamma^{r}_{\varphi\varphi},\nn\\
&\partial_r\Gamma^{\theta}_{r\theta},\;\partial_r\Gamma^{\varphi}_{tr},\;\partial_r\Gamma^{\varphi}_{r\varphi},\\
\label{Gc}&\partial_{\theta}\Gamma^{t}_{t\theta},\;\partial_{\theta}\Gamma^{t}_{\theta\varphi},\;\partial_{\theta}\Gamma^{r}_{r\theta},
\;\partial_{\theta}\Gamma^{\theta}_{tt},\;\partial_{\theta}\Gamma^{\theta}_{t\varphi},
\;\partial_{\theta}\Gamma^{\theta}_{rr},\;\partial_{\theta}\Gamma^{\theta}_{\theta\theta},\nn\\
&\partial_{\theta}\Gamma^{\theta}_{\varphi\varphi},\;\partial_{\theta}\Gamma^{\varphi}_{t\theta},\;\partial_{\theta}\Gamma^{\varphi}_{\theta\varphi},
\end{align}
and those obtained by symmetry.

\subsection{The electromagnetic field}
We assume the existence of a uniform magnetic field $B$ that is asymptotically parallel to the axis of symmetry (axis of rotation). The stress-energy tensor associated with $B$ is assumed to be small not to affect the geometry of the stationary, axisymmetric metric~\eqref{g1}. In empty space $F_{\mu\nu}$ is solution to $F^{\mu\nu}{}_{;\nu}=0$ the general expression of which may be postulated as
\begin{align}
\label{G1}&F_{tr}=F_1(r,\,\theta)+F_2(r,\,\theta)\cos2\theta,\\
\label{G2}&F_{r\varphi}=F_3(r,\,\theta)\sin^2\theta,\\
\label{G3}&F_{t\theta}=F_4(r,\,\theta)\sin2\theta,\;\; \partial_{\theta}F_4\propto \sin^n\theta\text{ or }0,\; (n\geq 1),\\
\label{G4}&F_{\theta\varphi}=F_5(r,\,\theta)\sin2\theta,\;\; \partial_{\theta}F_5\propto \sin^n\theta\text{ or }0,\; (n\geq 1),
\end{align}
where the five functions $F_1\to F_5$ and their $\theta$ derivatives are regular (finite) in empty space. The constraints, $\partial_{\theta}F_4\propto \sin^n\theta\text{ or }0$ and $\partial_{\theta}F_5\propto \sin^n\theta\text{ or }0$ with $n\geq 1$, ensure that the expressions of $F^{t\nu}{}_{;\nu}$ and $F^{\varphi\nu}{}_{;\nu}$ have finite limits as $\theta\to 0\text{ or }\pi$ (on the axis of rotation). Note that the ansatz~\eqref{G1} to~\eqref{G4} is conform to the different available expressions~\cite{Kerr1,vac,Wald,JP} for $F_{\mu\nu}$ due to the presence of a magnetic field $B$ asymptotically parallel to the axis of symmetry.

The ansatz~\eqref{G1} to~\eqref{G4} implies that the only \emph{nonvanishing} elements of $F_{\mu\nu}$, $\partial_r F_{\mu\nu}$, and $\partial_{\theta}F_{\mu\nu}$ \emph{on the $\theta=\pi/2$ plane} are
\begin{align}
\label{Fa}&F_{tr},\;F_{r\varphi},\\
\label{Fb}&\partial_r F_{tr},\;\partial_r F_{r\varphi},\\
\label{Fc}&\partial_{\theta}F_{tr},\;\partial_{\theta}F_{t\theta},\;\partial_{\theta}F_{\theta\varphi},
\end{align}
and those obtained by symmetry. These \emph{physical} properties~\eqref{Fa} to~\eqref{Fc} are generic as far as one is concerned with the metric~\eqref{g1} subject to~\eqref{g2}. They express the nonvanishing of some of the components of the Lorentz vector force acting on a test charged particle and of their directional derivatives in the $\theta=\pi/2$ plane.

\section{Perturbed circular motion\label{pcm}}
Given the metric~\eqref{g1}, along with the properties~\eqref{g2}, which, with the exception of the Taub--NUT rotating solutions, describes a large class of rotating singular and regular solutions. We assume the presence of a test magnetic filed the tensor $F_{\mu\nu}$ of which is of the form~\eqref{G1} to~\eqref{G4}. Using the properties~\eqref{Ga} to~\eqref{Gc} and~\eqref{Fa} to~\eqref{Fc} of these two fields we aim to decouple the set of equations governing the perturbed circular motion.

\subsection{Circular motion}
The equation,
\begin{equation}\label{p1}
\frac{\dd u^{\mu}}{\dd \tau}+\Gamma^{\mu}_{\alpha\beta}u^{\alpha}u^{\beta}=
e F^{\mu}_{\ \ \nu}u^{\nu},
\end{equation}
where $e\equiv q/m$, describes the motion of a particle of mass $m$, electric charge $q$, and four-velocity $u^{\mu}=\dd x^{\mu}/\dd \tau$. Here $F^{\mu}_{\ \ \nu}=g^{\mu\sigma}F_{\sigma\nu}$, the inverse metric $g^{\mu\sigma}$, and the connection $\Gamma^{\mu}_{\alpha\beta}$ are related to the unperturbed metric~\eqref{g1}.

For a circular motion in the equatorial plane ($\theta=\pi/2$), $u^{\mu}=(u^t,\,0,\,0,\,u^{\varphi})=u^t(1,\,0,\,0,\,\omega)$, where $\omega=\dd\varphi/\dd t$ is the angular velocity of the charged particle. The only equations describing such a motion are the $r$ component of~\eqref{p1} and the normalization condition $g_{\mu\nu}u^{\mu}u^{\nu}=-1$, which take the following forms, respectively
\begin{align}
\label{p2a}&\partial _r g_{tt} (u^t)^2+2 \partial _r g_{t\varphi}u^t u^{\varphi }+\partial _r g_{\varphi \varphi } (u^{\varphi })^2=-2 e (F_{rt} u^t+F_{r\varphi} u^{\varphi }),\\
\label{p2b}&g_{\text{tt}} (u^t)^2+2 g_{t\varphi} u^t u^{\varphi }
+g_{\varphi \varphi } (u^{\varphi })^2=-1,
\end{align}
where the metric and its derivatives are all evaluated at $\theta=\pi/2$. These equations can in principal be solved for ($u^t,\,u^{\varphi }$) in terms of the radius of the circle. One may also solve for ($u^t,\,\omega$) as follows:
\begin{align}
\label{p3a}&(u^t)^2=-\frac{1}{g_{tt}+2 g_{t\varphi } \omega +g_{\varphi \varphi } \omega ^2},\\
\label{p3b}&\frac{\partial _r g_{tt}+2 \partial _r g_{t\varphi } \omega +\partial _r g_{\varphi \varphi } \omega ^2}{\sqrt{-(g_{tt}+2g_{t\varphi } \omega +g_{\varphi \varphi } \omega ^2)}}=-2 e (F_{rt}+F_{r\varphi } \omega ).
\end{align}

\subsection{ECOs\label{ECOs}}
We assume that the charged particle has been displaced from its circular position $x^{\mu}$ by an infinitesimal amount $\eta^{\mu}$ so that its actual position is now $X^{\mu}=x^{\mu}+\eta^{\mu}$ and its 4-velocity is $U^{\mu}=u^{\mu}+\dot{\eta}^{\mu}$ ($~\dot{}\equiv \dd /\dd\tau$) with $u^{\mu}$ being a solution to~\eqref{p2a} and~\eqref{p2b} [where only ($u^t,\,u^{\varphi}$) are nonzero]. Using this in
\begin{equation}
\frac{\dd U^{\mu}}{\dd \tau}+\Gamma^{\mu}_{\alpha\beta}(X^{\sigma})U^{\alpha}U^{\beta}=
e F^{\mu}_{\ \ \nu}(X^{\sigma})U^{\nu},
\end{equation}
along with~\eqref{p1} and keeping only linear terms in $\eta$ we arrive at~\cite{Kerr1}\footnote{In Eq. (9) of Ref~\cite{Kerr1}, $\bar{F}^{\mu}_{\ \ \nu}\xi^{\nu}$ should read $\bar{F}^{\mu}_{\ \ \nu}\dot{\xi}^{\nu}$.}
\begin{equation}\label{p4}
\ddot{\eta}^{\mu}+2\Gamma^{\mu}_{\alpha\beta}u^{\alpha}\dot{\eta}^{\beta}
+\partial_{\nu}\Gamma^{\mu}_{\alpha\beta}u^{\alpha}u^{\beta}\eta^{\nu}
=e(F^{\mu}_{\ \ \nu}\dot{\eta}^{\nu}+\partial_{\lambda}F^{\mu}_{\ \ \nu}u^{\nu}\eta^{\lambda}),
\end{equation}
where $\Gamma^{\mu}_{\alpha\beta}$, $F^{\mu}_{\ \ \nu}$, and their derivatives are evaluated at $\theta=\pi/2$.

\subsubsection{The $\theta$ component}
The properties~\eqref{Fa} to~\eqref{Fc}, imply that
\begin{equation}
F^{\theta}_{\ \ \nu}=g^{\theta\theta}F_{\theta\nu},\;\partial_rF^{\theta}_{\ \ t},\;\partial_rF^{\theta}_{\ \ \varphi},
\end{equation}
vanish at $\theta=\pi/2$. Using this along with~\eqref{Ga} to~\eqref{Gc} in~\eqref{p4} the $\theta$ component decouples and takes the form of an oscillating vertical motion (perpendicular to the $\theta=\pi/2$ plane)
\begin{align}
\label{p5a}&\ddot{\eta}^{\theta}+\omega_{\theta}^2\eta^{\theta}=0,\\
\label{p5b}&\omega_{\theta}^2\equiv \partial_{\theta}\Gamma^{\theta}_{tt}(u^t)^2+2\partial_{\theta}\Gamma^{\theta}_{t\varphi}u^tu^{\varphi}
+\partial_{\theta}\Gamma^{\theta}_{\varphi\varphi}(u^{\varphi})^2\nn\\
&\qquad +e[\partial_{\theta}(g^{\theta\theta}F_{t\theta})u^{t}+\partial_{\theta}(g^{\theta\theta}F_{\varphi\theta})u^{\varphi}],\\
\label{p5c}&\quad\, =\partial_{\theta}(\Gamma^{\theta}_{ij}u^iu^j+eg^{\theta\theta}F_{i\theta}u^i),\qquad (i,\,j=t,\,\varphi),
\end{align}
where the derivatives with respect to $\theta$ are evaluated at $\theta=\pi/2$. In the compact expression~\eqref{p5c} it is understood that ($u^t,\,u^{\varphi}$) do not depend on $\theta$. Since ($u^t,\,u^{\varphi}$) are solutions to~\eqref{p2a} and~\eqref{p2b}, $\omega_{\theta}^2$ depends on the detailed circular motion and on the electromagnetic field. The stability of the circular motion is partially guaranteed by the positiveness of $\omega_{\theta}^2$.

\subsubsection{The $t$ and $\varphi$ components}
The properties~\eqref{Ga} to~\eqref{Gc} and~\eqref{Fa} to~\eqref{Fc} will allow us to decouple the $r$ equation of~\eqref{p4}; however, this can be done in two steps. First, we need expressions for $\dot{\eta}^{t}$ and $\dot{\eta}^{\varphi}$ which are obtained from the $t$ and $\varphi$ components of~\eqref{p4} by integration
\begin{equation}\label{p6}
\dot{\eta}^{j}=(-2\Gamma^{j}_{ri}u^i+eg^{ji}F_{ir})\eta^r,\qquad (i,\,j=t,\,\varphi),
\end{equation}
where we have set the constants of integration to zero.

\subsubsection{The $r$ component}
Using~\eqref{Ga} to~\eqref{Gc} and~\eqref{Fa} to~\eqref{Fc}, the $r$ equation of~\eqref{p4} is first brought to the form
\begin{multline}\label{p7}
\ddot{\eta}^{r}+\partial_{r}(\Gamma^{r}_{ij}u^iu^j+eg^{rr}F_{ir}u^i)\eta^{r}
+(2\Gamma^{r}_{ji}u^i+eg^{rr}F_{jr})\dot{\eta}^{j}=0,\\
\qquad (i,\,j=t,\,\varphi),
\end{multline}
where it is understood that ($u^t,\,u^{\varphi}$) are not perturbed: $\partial_{r}(eg^{rr}F_{ir}u^i)=e\partial_{r}(g^{rr}F_{ir})u^i$. Then, using~\eqref{p6} to eliminate $\dot{\eta}^{j}$, we decouple it into an equation describing an oscillating radial motion,
\begin{equation}\label{p8a}
\ddot{\eta}^{r}+\omega_{r}^2\eta^{r}=0,
\end{equation}
with frequency $\omega_{r}$ given by
\begin{multline}\label{p8b}
\omega_{r}^2\equiv \partial_{r}(\Gamma^{r}_{ij}u^iu^j+eg^{rr}F_{ir}u^i)+2e(g^{ik}\Gamma^{r}_{kj}-g^{rr}\Gamma^{i}_{rj})F_{ir}u^j\\
+e^2g^{rr}g^{ij}F_{ir}F_{jr}-4\Gamma^{r}_{ik}\Gamma^{k}_{rj}u^iu^j,\qquad (i,\,j,\,k=t,\,\varphi).
\end{multline}
The stability of the oscillating radial motion is ensured by the positiveness of $\omega_{r}^2$. The circular motion is considered stable if both local frequencies $\omega_{r}^2$ and $\omega_{\theta}^2$ are positive.

In the limit of small magnetic field, the decoupling of the perturbed circular motion for the Kerr black hole has been achieved in~\cite{Kerr1,Kerr2}.

In summarizing, provided the background metric is given by~\eqref{g1}, we have relied on the properties~\eqref{Ga} to~\eqref{Gc} and~\eqref{Fa} to~\eqref{Fc} to obtain the decoupled equations for $\eta^r$ and $\eta^{\theta}$. In concluding, the perturbed circular motion of a charged particle, in the presence of a \emph{relatively weak} magnetic field parallel to the axis of rotation, obeys all the equations derived in this section \emph{irrespective of the value of the rotation parameter} within its limits of variation. \emph{Nonnegligible} magnetic fields would violate the circularity of spacetime~\cite{vio1,vio2} which, in turn, would generate other cross terms in the metric than $\dd t\dd\vp$ (see footnote~\ref{foot1}).

\subsection{Relation of $\omega_{r}$ to the effective potential function}
The general motion of (un)charged particles in the geometry of a stationary solution immersed in an external magnetic field is not separable. If restricted to the $\theta=\pi/2$ plane, the motion becomes separable where the radial motion is usually brought, after a first integration, to some first order differential equation including a potential function. The choice of the potential function is at the author's stylistic choice and discretion. Among these choices one finds $\dot{r}^2=V(r)$, $\dot{r}^2=\mathcal{R}(r)/r^4$, or $\dot{r}^2=R(r)/r^3$. Assume that the radial motion has been brought to
\begin{equation}\label{f1}
\dot{r}^2=V(r),
\end{equation}
where $V(r)$ is the potential function depending on some constants of the motion. Differentiating~\eqref{f1} with respect to $\tau$ we obtain
\begin{equation}\label{f2}
\ddot{r}=\frac{1}{2}~V'(r),
\end{equation}
where we have dropped $\dot{r}$ from both sides. Here $V'\equiv \partial_r V$. Assume that the radial motion has been displaced from its stable path $r_0(\tau)$ by the small amount $\eta^r$: $r=r_0+\eta^r$. Using this in~\eqref{f2} along with $V'(r)=V'(r_0)+V''(r_0)\eta^r+\cdots$ and $\ddot{r}_0=\frac{1}{2}~V'(r_0)$, we obtain
\begin{equation}\label{f3}
\ddot{\eta}^{r}-\frac{V''(r_0)}{2}~\eta^{r}=0.
\end{equation}
On comparing this with~\eqref{p8a}, we express $\omega_{r}^2$ in terms of the second derivative of the potential function
\begin{equation}\label{f4}
\omega_{r}^2(r_0)=-\frac{V''(r_0)}{2}.
\end{equation}
A similar formula has been derived in~\cite{epi}. Thus, in the representation~\eqref{f1} the second derivative of the potential at $r_0$ should be negative to have a stable path there: $V''(r_0)<0$. If the motion is circular, $r_0$ is a constant and $V(r_0)=V'(r_0)=0$, so the condition $V''(r_0)<0$ implies that the potential should have a maximum value at $r_0$ to ensure stability of the circular orbit.

At the inner stable circular orbit (isco), or marginally stable circular orbit, $V''(r_{\text{isco}})=0$ and this consists a limiting case of stable circular orbits. This extra condition yields
\begin{equation}\label{f5}
\omega_{r}^2(r_{\text{isco}})=0.
\end{equation}

\section{Properties of the perturbed circular motion\label{ppcm}}
The small quantities $\eta^{\mu}$ are obtained by integrating the above-determined equations~\eqref{p5a}, \eqref{p6}, and~\eqref{p8a}
\begin{align}\label{p9}
&\eta^r=C_r\cos\omega_r \tau,\quad \eta^{\theta}=\hat{C}_{\theta}\cos\omega_{\theta} \tau +\tilde{C}_{\theta}\sin\omega_{\theta} \tau,\\
&\eta^t=(-2\Gamma^{t}_{ri}u^i+eg^{ti}F_{ir})C_r~\frac{\sin\omega_r \tau}{\omega_r}+C_t,\\ &\eta^{\varphi}=(-2\Gamma^{\varphi}_{ri}u^i+eg^{\varphi i}F_{i r})C_r~\frac{\sin\omega_r \tau}{\omega_r}+C_{\varphi},
\end{align}
where we have selected some of the integration constants and the remaining constants, ($C_t,\,C_r,\,\hat{C}_{\theta},\,\tilde{C}_{\theta},\,C_{\varphi}$), are considered small; ($C_t,\,C_{\varphi}$) could be set equal to 0.

The frequencies ($\omega_{r},\,\omega_{\theta}$) are generally different and are different from the ZAMO's angular velocity $\Omega\equiv -g_{t\varphi}/g_{\varphi\varphi}$~\eqref{g1}, the Keplerian or orbital frequency $\omega_{\text{K}}\equiv u^{\varphi}=\omega u^t$~\eqref{p3b}, and the Larmor angular frequency $\omega_{\text{L}}\equiv eB$. 

We cannot proceed further until a metric is prescribed, however, earlier investigations for the Kerr-black-hole perturbed circular motion~\cite{Tursunov} have proven the instability of the vertical motion ($\omega_{\theta}^2<0$) of the perturbed charged-particle circular motion in some regions where both the radial and vertical motions of the perturbed Kerr circular geodesics~\footnote{By {\textquotedblleft Kerr circular geodesics\textquotedblright} we mean circular time-like orbits of the Kerr black hole.} are stable. This shows that the usually made hypothesis of a plane accretion disk is bold if the disk contains charged particles plunged into a cosmic magnetic field since, in this case, one may have $\omega_{\theta}^2<0$ resulting in an unbounded or chaotic vertical motion violating the hypothesis of plane accretion.

It has been noticed too~\cite{Tursunov} that, in the case of the anti-Larmor orbits (corresponding to a repulsive magnetic force), the order relation $\omega_{r}\simeq \omega_{\theta}\gg \omega_{\text{K}}$ holds resulting in toroidal orbits with almost transverse circular sections.

The ZAMO's angular velocity $\Omega$ plays no role in such investigations.

\section{Application\label{secapp}}
In order to provide an application of the formulas derived in Sec.~\ref{pcm} we need to fix a metric and the corresponding electromagnetic field when the spacetime is immersed in a uniform magnetic field.

\subsection{The metric and electromagnetic field}
For the general metric~\eqref{g1} it does not seem to be there an available expression for $F_{\mu\nu}$ when the spacetime is immersed in a uniform magnetic field. We need to restrict ourselves to the following rotating metric,
\begin{align}\label{m1}
\dd s^2 =&-\Big(1-\frac{2f}{\ro^2}\Big)\dd t^2+\frac{\ro^2}{\De}\,\dd r^2
-\frac{4af \sin ^2\theta}{\ro^2}\,\dd t\dd \vp\nn\\
\quad & +\ro^2\dd \ta^2+\frac{\Si\sin ^2\theta}{\ro^2}\,\dd \vp^2,
\end{align}
where
\begin{align}\label{m2}
&\ro^2=r^2+a^2\cos^2\ta,\quad \De(r) = r^2-2f(r)+a^2, \nn\\
&\Si =(r^2+a^2)^2-a^2\De\sin^2\ta,\\
&a:\text{ rotation parameter},\nn
\end{align}
which was derived in~\cite{GG} then rederived in~\cite{gen1}. This metric is a special case of~\eqref{g1} but it is almost as general as the metric~\eqref{g1} is. It describes a variety of solutions including (A) Schwarzschild ($a=0$), Reissner--Nordstr\"om ($a=0$), Kerr, Kerr--Newman metrics, the Schwarzschild--MOG ($a=0$) and Kerr--MOG black holes of the modified gravity (MOG)~\cite{MOG,MOG2}, and their trivial generalizations the Reissner--Nordstr\"om--MOG ($a=0$) and Kerr--Newman--MOG, and some phantom Einstein--Maxwell--dilaton black holes~\cite{phantom,multi}. It also includes (B) nonrotating regular black holes~\cite{regular1}-\cite{azth} and their rotating counterparts~\cite{gen1,gen2,gen3} as well as some nonrotating black holes of $F(R)$ and $F(T)$ gravities~\cite{m3,T}, and (C) noncommutative and quantum-corrected black holes~\cite{SS3}-\cite{qcbh4}. Rotating wormholes~\cite{az3}-\cite{az6} are also described by~(\ref{m1}, \ref{m2}). In fact, the list of rotating solutions described by~(\ref{m1}, \ref{m2}) is still open: Recently some workers~\cite{s1,s2,s3,s4,s5,s6,s7,s10} used our porcedure~\cite{gen1} to derive rotating solution of the form~(\ref{m1}, \ref{m2}).

With the expressions $N=(g_{t\varphi}^2-g_{tt}g_{\varphi\varphi})/g_{\varphi\varphi}$, $A=g_{\theta\theta}/r^2$, $D=g_{\varphi\varphi}/(r^2\sin^2\theta)$, and $\omega=-g_{t\varphi}/g_{\varphi\varphi}$ it is easy to check that, for all functions $f(r)$, the stationary, axisymmetric metric~(\ref{m1}, \ref{m2}) satisfies the properties~\eqref{Ga} to~\eqref{Gc}.

For the class (A) of singular black holes the function $f(r)$ in~(\ref{m1}, \ref{m2}) is of the form,
\begin{equation}\label{f}
    f(r)=f_1r+f_2,
\end{equation}
where the constants ($f_1,\,f_2$) do not depend on the rotation parameter $a$. Table 1 of Ref.~\cite{vac} provides the values of ($f_1,\,f_2$) for the set (A) of singular black hole solutions. For instance, $f_1=M$ and $f_2=-Q^2/2$ for the Reissner--Nordstr\"om and Kerr--Newman black holes and $f_1=(1+\ka^2)M$ and $f_2=-(1+\ka^2)(\ka^2M^2+Q^2)/2$ for the Kerr--Newman--MOG black hole. Here $\kappa$ is the universal ratio of the scalar charge $Q_s$ to the mass $M$ of the black hole~\cite{MOG,MOG2}.

We assume the existence of a uniform \emph{test} magnetic field $B$ that is asymptotically parallel to the axis of symmetry. The stress-energy tensor associated with $B$ is assumed to be small not to affect the geometry of the stationary, axisymmetric metric~(\ref{m1}, \ref{m2}). If the metric~(\ref{m1}, \ref{m2}) represents a charged \emph{singular} black hole, with central charge $Q$, the total vector potential $A^{\mu}$ and the electromagnetic tensor $F_{\mu\nu}=\partial_{\mu}A_{\nu}-\partial_{\nu}A_{\mu}$ for \emph{slowly} rotating black holes are given by~\cite{vac}
\begin{multline}\label{ef1}
A^{\mu}=\Big[\frac{Qr}{r^2-2f}+aB\Big(1+\frac{f_2\sin^2\ta}{r^2}\Big)\Big]\xi^{\ \mu}_{t}\\
+\Big[\frac{B}{2}\Big(1+\frac{2f_2}{r^2}\Big)+\frac{Qa}{r(r^2-2f)}\Big]\xi^{\ \mu}_{\vp},
\end{multline}
\begin{align}\label{ef2}
&F_{tr}=-\frac{Q}{r^2}+\frac{B a [4 f_2+3 f_1 r+(4 f_2+f_1 r) \cos  2 \theta ]}{2 r^3},\nn\\
&F_{t\theta}=\frac{B a (2 f_2+f_1 r) \sin  2 \theta }{r^2},\nn\\
&F_{r\vp}=B r \sin ^2 \theta -\frac{Q a \sin ^2 \theta }{r^2},\\
&F_{\theta \varphi }=\frac{B }{2}(2 f_2+r^2) \sin  2 \theta +\frac{Q a \sin  2 \theta }{r},\nn
\end{align}
and the nonvanishing expressions of $F^{\mu\nu}$ are given in the Appendix. These expressions correspond to a magnetic field oriented in the positive direction along the axis of rotation~\footnote{There was a sign slip in~\cite{vac} where the expressions of $F_{\mu\nu}$ correspond to a magnetic field oriented in the negative direction along the axis of rotation. Related statements and formulas should be changed correspondingly.}.

It is straightforward to check that $F_{\mu\nu}$ satisfies the properties~\eqref{Fa} to~\eqref{Fc}.

The r.h.s of~\eqref{ef1} includes all linear terms in $B$, thus adding corrections to Wald's formula~\cite{Wald}:
\begin{equation}\label{ef4}
aB\xi^{\ \mu}_{t}
+\frac{B}{2}\xi^{\ \mu}_{\vp}.
\end{equation}
As has been emphasized in~\cite{vac}, Wald's formula applies only to the Schwarzschild and Kerr black holes where $f(r)=f_1r=Mr$ ($f_1 = M$ and $f_2\equiv 0$). The terms proportional to $Q$ in~\eqref{ef1} are just the linear parts with respect to $a$ of
\begin{equation}\label{ef3}
\frac{Qr(r^2+a^2)}{\rho^2\Delta}~\xi^{\ \mu}_{t}+\frac{Qra}{\rho^2\Delta}~\xi^{\ \mu}_{\vp},
\end{equation}
which yields the one-form $A_{\mu}\dd x^{\mu}=-Qr(\dd t-a\sin^2\ta\dd \vp)/\ro^2$ for charged black holes. We see that the general expression~\eqref{ef1} is not just a mere linear combination of~\eqref{ef4} and~\eqref{ef3}; rather, there are ``interaction terms" proportional to $aBf_2$ and to $Bf_2$. The former interaction term includes the effects of rotation, the magnetic field, and all that expressed in $f_2$ ($f_2$ may depend on the electric charge $Q$, the MOG scalar charge, and possibly other parameters~\cite{vac}), and the latter interaction term includes only the effects of the magnetic field and all that expressed in $f_2$.

If applied to \emph{charged} solutions, Wald's formula results in nonvanishing currents ($J^t,\,J^{\varphi}$) in empty regions of space~\cite{vac} (where $J^{\mu}=F^{\mu\nu}{}_{;\nu}=(\sqrt{|g|}F^{\mu\nu}){}_{,\nu}/\sqrt{|g|}$). The purpose of the extra correction terms given in~\eqref{ef1} is to make these currents vanish identically in empty regions of space. Moreover, these correction terms introduce extra terms to the Lorentz force. While the currents ($J^t,\,J^{\varphi}$), generated by Wald's formula, vanish asymptotically (as $r\to\infty$), his formula still cannot be applied to charged, static or rotating, solutions where in most cases one is interested in accretion phenomena that take place in the vicinity of the innermost stable circular orbit (isco) whose radius is of the order of the event horizon.

Another expression for $F_{\mu\nu}$ for \emph{slowly} rotating uncharged black holes has been provided in Ref.~\cite{Kerr1} the linear terms of which, with respect to $a$, reduce to the remaining expression of~\eqref{ef2} after setting $Q=0$, $f_1=M$, and $f_2=0$ (up to a global sign due to the fact that in Ref.~\cite{Kerr1} the definition $F_{\mu\nu}=\partial_{\nu}A_{\mu}-\partial_{\mu}A_{\nu}$ has been used). We do not know whether Aliev--Galtsov expression for $F_{\mu\nu}$ remains valid up to $a^2$ terms. The leading terms of the Petterson expression for $F_{\mu\nu}$~\cite{JP}, which applies to the Kerr black hole only, also reduces to~\eqref{ef2} upon an appropriate choice of parameters~\cite{vac}.

\subsection{The weakly magnetized Kerr--Newman black hole}
The metric of the Kerr--Newman black hole is of the form~(\ref{m1}, \ref{m2}) with $f=f_1r+f_2=Mr-Q^2/2$. The usually used parameters for such investigations~\cite{Kerr1,Kerr2,Kerr3} are the so-called magnetic field parameter
\begin{equation}\label{w1}
b=\frac{eB}{2},
\end{equation}
($e\equiv q/m$) and the dimensionless ratios
\begin{equation}\label{w2}
\mathcal{B}=\frac{eBMG}{c^4},\qquad \mathcal{Q}=\frac{eQ}{4\pi\epsilon_0GM},
\end{equation}
where we have introduced the known constants ($c,\,G,1/(4\pi\epsilon_0)$). It may be agreed upon that the cosmic magnetic field $B$ is weak of the order of $10^4-10^8$ Gauss but $\mathcal{B}$, which is a measure of the Lorentz force relative to the gravitational force, is large enough~\cite{Kerr3} to have astrophysical observational effects~\cite{Kerr2}.

\begin{figure*}
\centering
\includegraphics[width=0.33\textwidth]{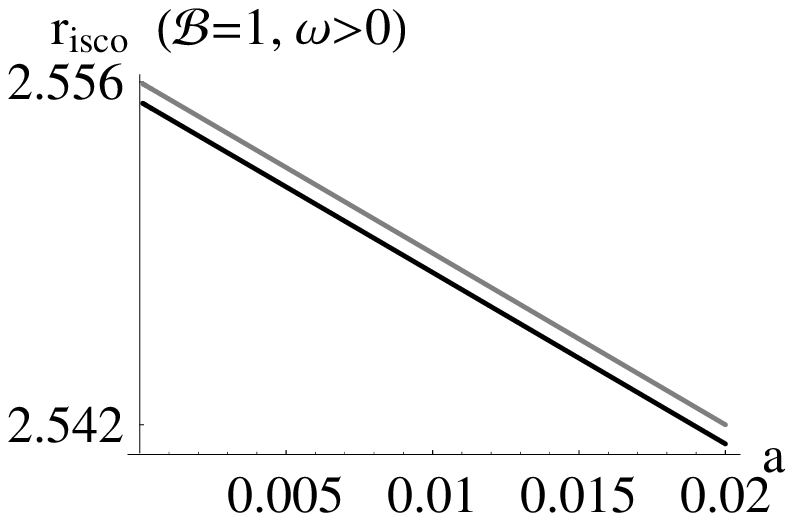} \includegraphics[width=0.33\textwidth]{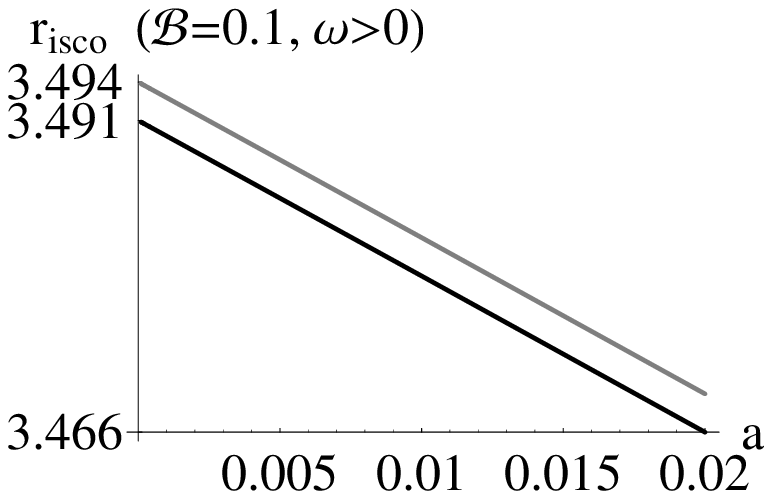} \includegraphics[width=0.33\textwidth]{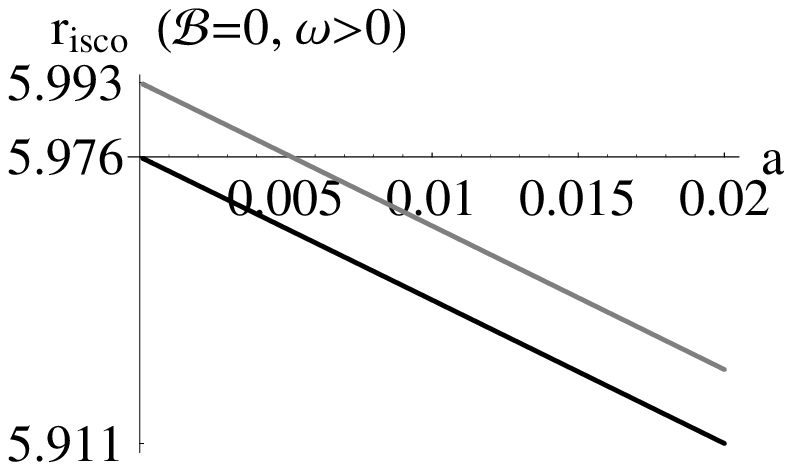} \\
\includegraphics[width=0.33\textwidth]{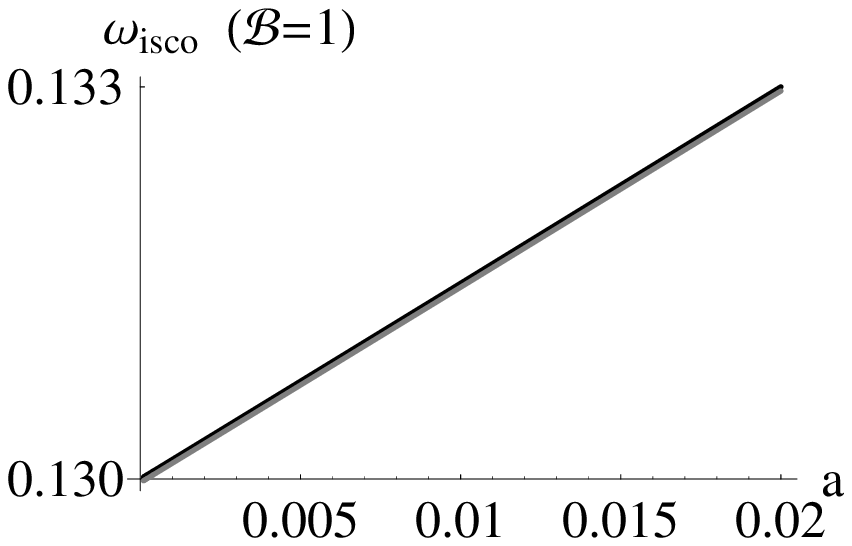} \includegraphics[width=0.33\textwidth]{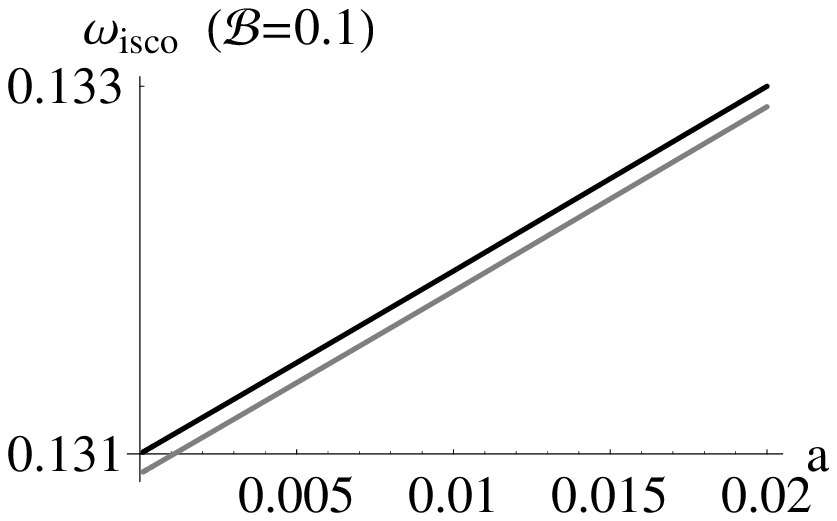} \includegraphics[width=0.33\textwidth]{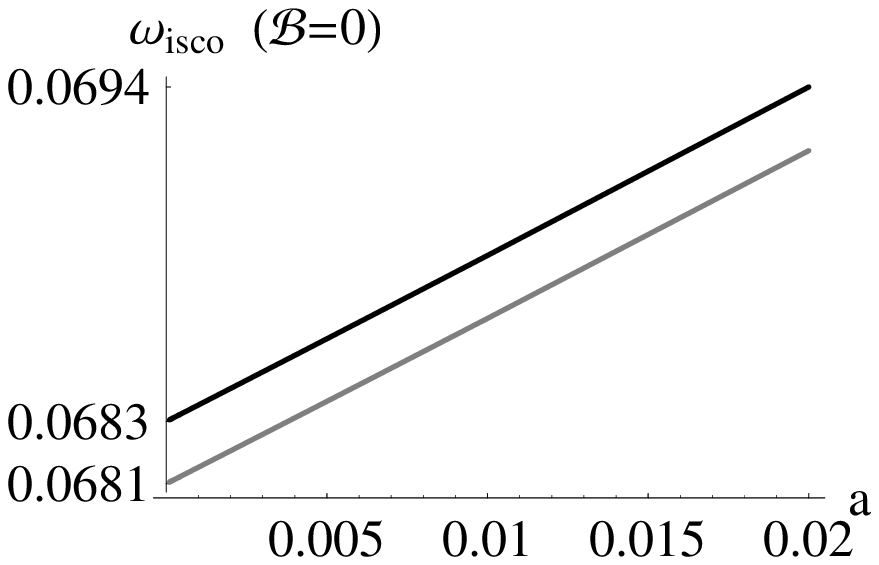} \\
\caption{\footnotesize{Plots of $r_{\text{isco}}$ and $\omega_{\text{isco}}$ ($=u^{\varphi}/u^t$) in terms of $a$ for $M=1$ (mass of the Kerr--Newman BH), $Q=1/10$ (charge of the Kerr--Newman BH), $\mathcal{Q}\equiv eQ/M=1/500$ for the black plots, and $Q=-1/10$, $\mathcal{Q}=-1/500$ for the grey plots. The case of prograde circular orbits ($u^{\varphi}>0$) of a charged particle with mass $m$ and charge $q=em$. The plots show the effect of the magnetic field $B$ encoded in the dimensionless ratio $\mathcal{B}$~\eqref{w2}, which is proportional to $eBM$. These solutions have been determined upon solving numerically Eqs.~\eqref{w3}, \eqref{w4}, and~\eqref{f5}, with $\omega_{r}^2$ being given by~\eqref{p8b}.}}\label{Fig1}
\end{figure*}

\begin{figure*}
\centering
\includegraphics[width=0.33\textwidth]{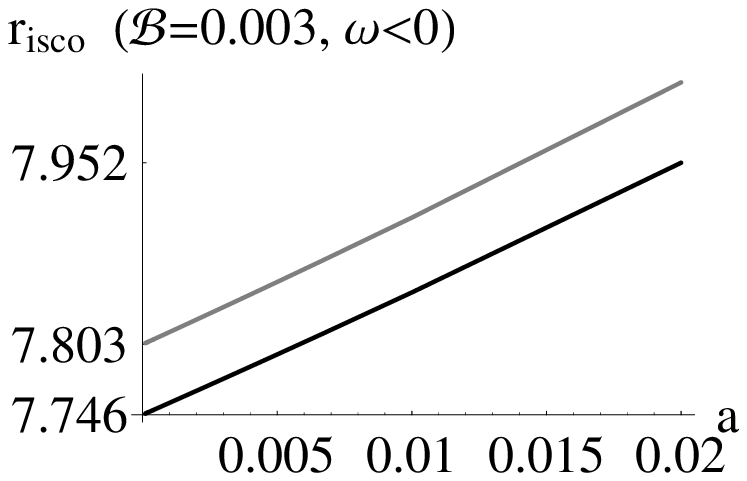} \includegraphics[width=0.33\textwidth]{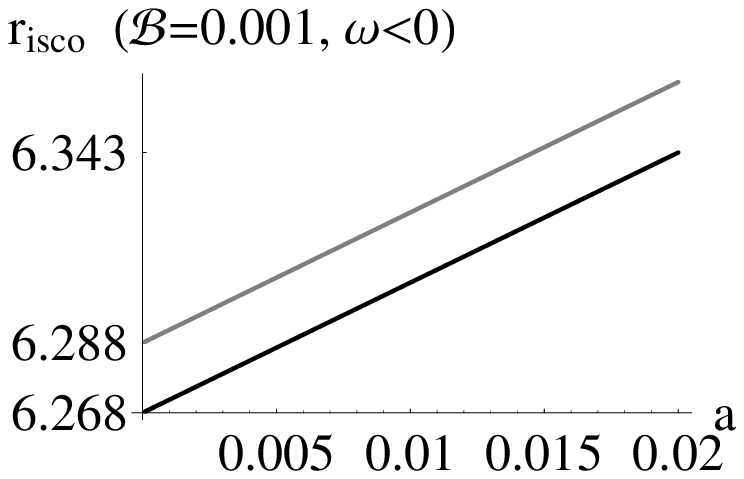} \includegraphics[width=0.33\textwidth]{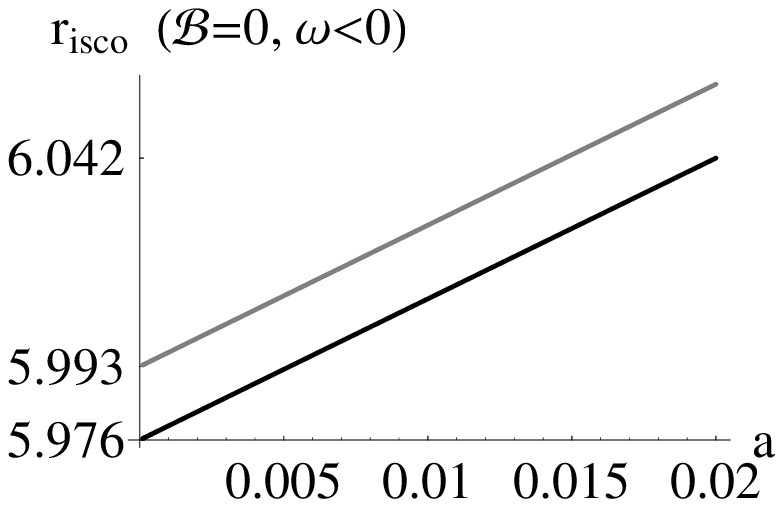} \\
\includegraphics[width=0.33\textwidth]{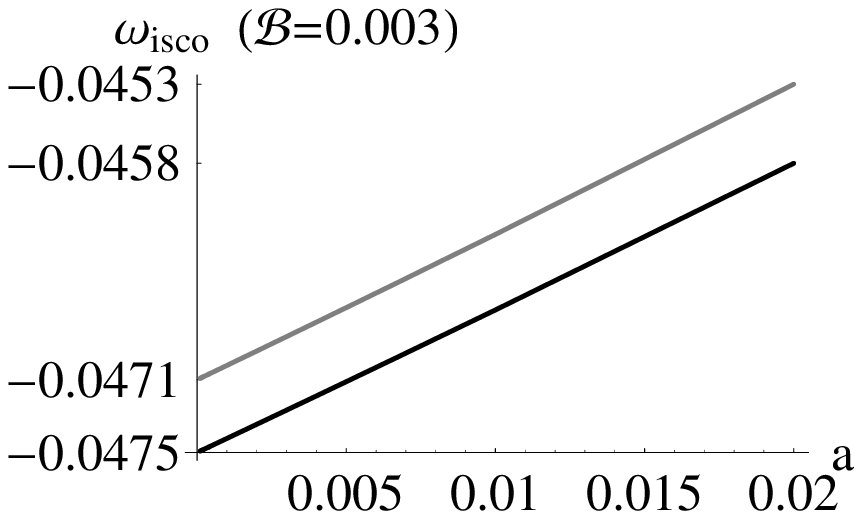} \includegraphics[width=0.33\textwidth]{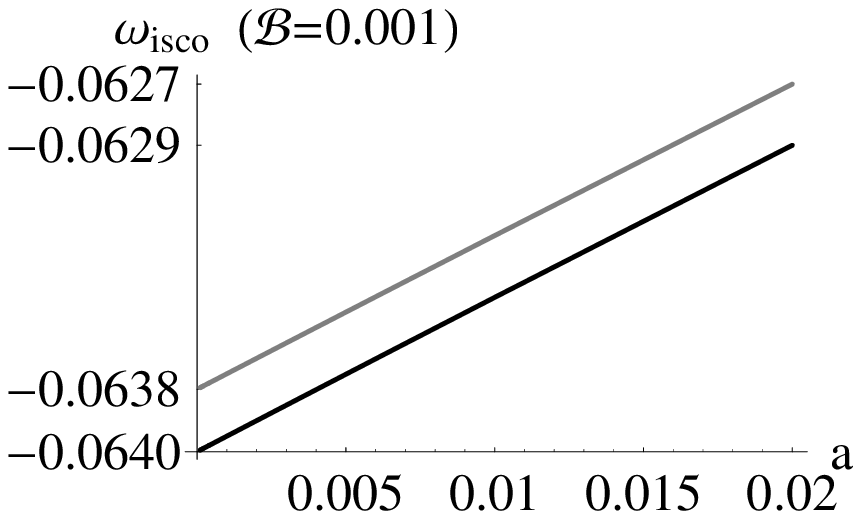} \includegraphics[width=0.33\textwidth]{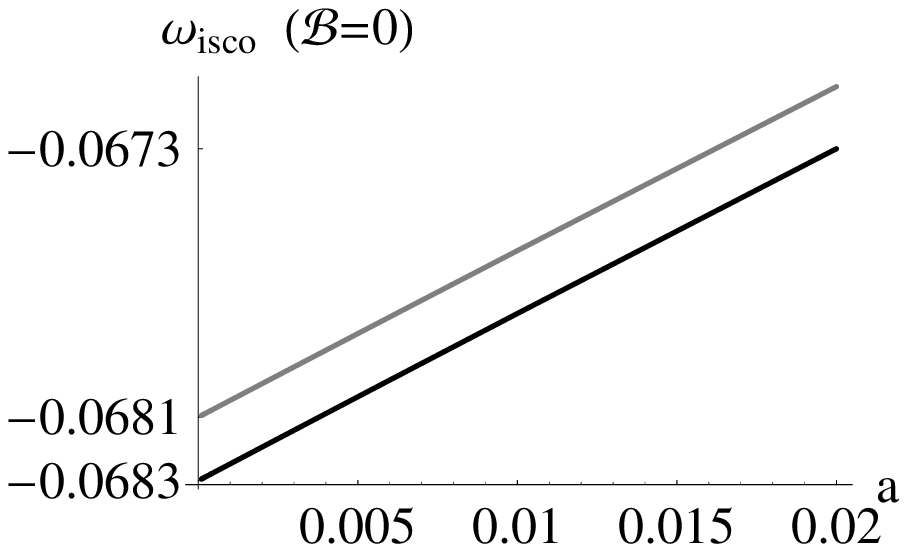} \\
\caption{\footnotesize{Plots of $r_{\text{isco}}$ and $\omega_{\text{isco}}$ ($=u^{\varphi}/u^t$) in terms of $a$ for $M=1$ (mass of the Kerr--Newman BH), $Q=1/10$ (charge of the Kerr--Newman BH), $\mathcal{Q}\equiv eQ/M=1/500$ for the black plots, and $Q=-1/10$, $\mathcal{Q}=-1/500$ for the grey plots. The case of retrograde circular orbits ($u^{\varphi}<0$) of a charged particle with mass $m$ and charge $q=em$. The plots show the effect of the magnetic field $B$ encoded in the dimensionless ratio $\mathcal{B}$~\eqref{w2}, which is proportional to $eBM$. These solutions have been determined upon solving numerically Eqs.~\eqref{w3}, \eqref{w4}, and~\eqref{f5}, with $\omega_{r}^2$ being given by~\eqref{p8b}. \emph{Radially stable prograde circular orbits exit for the whole range of $\mathcal{B}$~\eqref{w4}, while for retrograde orbits it seems that there is some critical value $\mathcal{B}_c$ beyond which no radially stable retrograde circular orbit exists (in the vicinity of $r_{\text{isco}}$ of the prograde circular orbits); however, as we shall see in Sec.~\ref{secqpos}, vertically stable circular orbits do exist (in the vicinity of $r_{\text{isco}}$ of the prograde circular orbits). For $\mathcal{Q}=+1/500$~\eqref{w4} we find $\mathcal{B}_c=0.003204$ and for $\mathcal{Q}=-1/500$ we find $\mathcal{B}_c=0.003189$. That's why we have chosen $\mathcal{B}\leq 0.003$ in this figure.} This statement, which we will improve in Sec.~\ref{seccon}, remains valid in the vicinity of $r_{\text{isco}}$.}}\label{Fig2}
\end{figure*}

\subsubsection{Circular orbits of the Kerr--Newman black hole}
Circular motions of un(charged) particles in the geometry of the Kerr--Newman black holes and its extensions including a Taub-NUT parameter or a cosmological constant have been investigated in a couple of references~\cite{Kerr1} and~\cite{cr1}-\cite{cr5} but none of the existing references includes effects of a magnetic field on the circular motion of charged particles. The equations we have derived so far will allow us to tackle this problem. In such investigations one is mostly interested in the isco positions that are determined upon solving the set of equations~\eqref{p2a}, \eqref{p2b}, and~\eqref{f5}. These isco positions determine the loci of accretion disks where radiation phenomena occur.

\begin{figure*}
	\centering
	\includegraphics[width=0.33\textwidth]{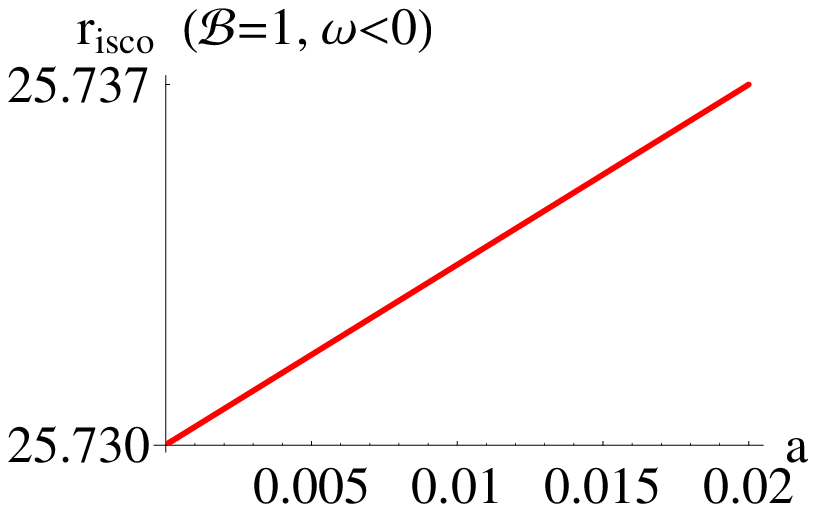} \includegraphics[width=0.33\textwidth]{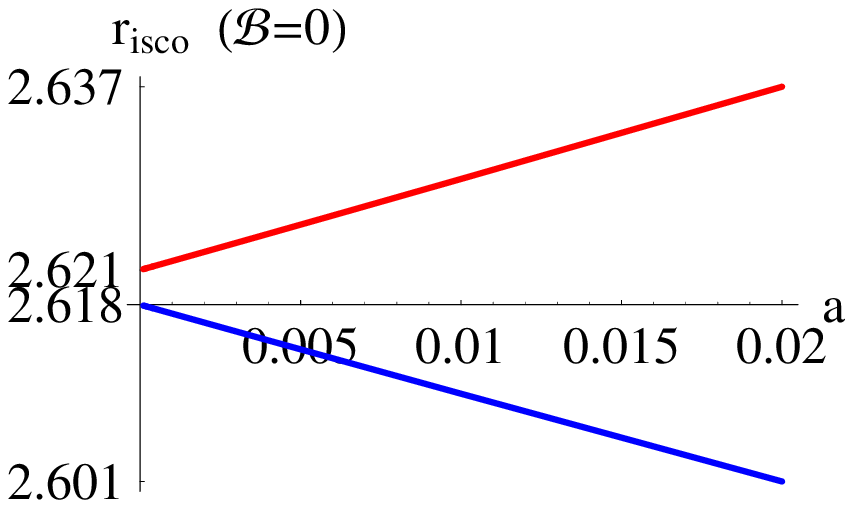} \\
	\includegraphics[width=0.33\textwidth]{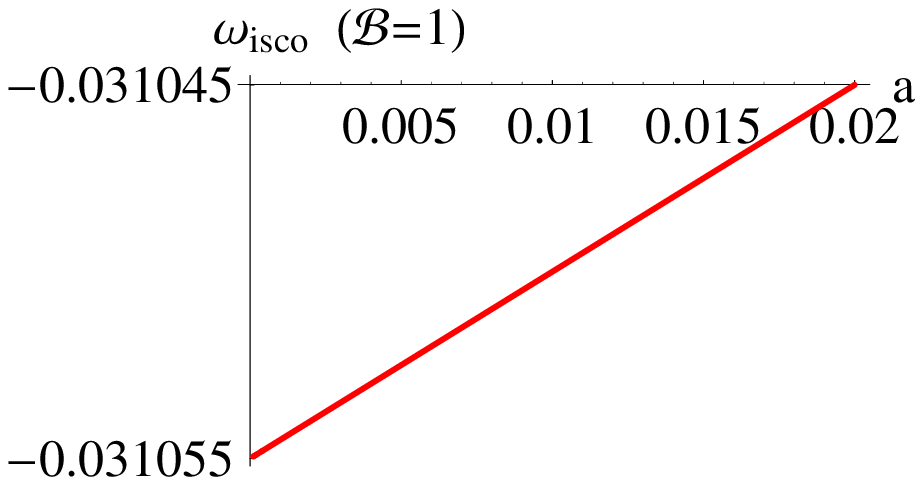} \includegraphics[width=0.33\textwidth]{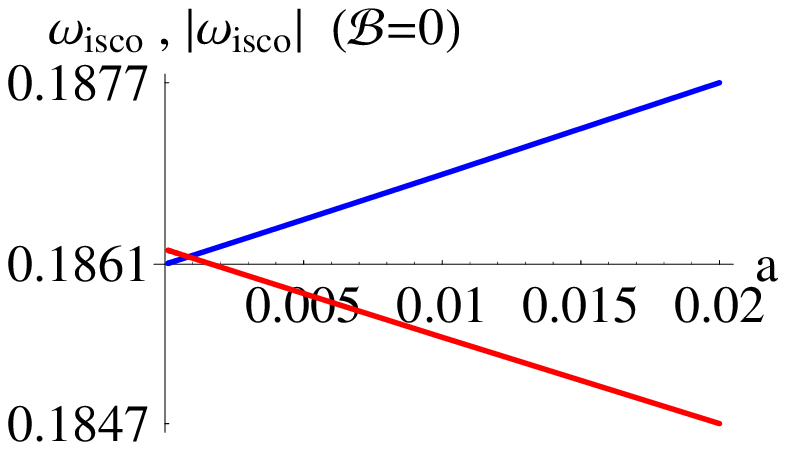} \\
	\caption{\footnotesize{Plots of $r_{\text{isco}}$ and $\omega_{\text{isco}}$ ($=u^{\varphi}/u^t$) in terms of $a$ for $M=1$ (mass of the Kerr--Newman BH), $Q=1/10$ (charge of the Kerr--Newman BH), and $\mathcal{Q}\equiv eQ/M=500$. The blue plots represent prograde circular orbits ($u^{\varphi}>0$) and the red plots represent retrograde circular orbits ($u^{\varphi}<0$) of a charged particle with mass $m$ and charge $q=em$. The plots show the effect of the magnetic field $B$ encoded in the dimensionless ratio $\mathcal{B}$~\eqref{w2}, which is proportional to $eBM$. For $\mathcal{B}=1$ there are no prograde circular orbits (for the values of the parameters used in this figure, prograde circular orbits do not exist for $\mathcal{B}>\bar{\mathcal{B}}_c=0.110000$). For $\mathcal{B}=0$ both prograde and retrograde circular orbits exist. In the lower-right plot we have represented $|\omega_{\text{isco}}|$ for the retrograde circular orbits.  These solutions have been determined upon solving numerically Eqs.~\eqref{w3}, \eqref{w4}, and~\eqref{f5}, with $\omega_{r}^2$ being given by~\eqref{p8b}.}}\label{Fig3}
\end{figure*}

It is out of question to deal analytically with Eqs.~\eqref{p2a}, \eqref{p2b}, and~\eqref{f5} nor is it welcome to write down the full expression of Eq.~\eqref{f5} in the Kerr--Newman geometry with $\omega_{r}^2$ being given by~\eqref{p8b}. Equations~\eqref{p2a} and \eqref{p2b} take rather simple expressions:
\begin{align}
\label{w3}&Q^2 (u^t-a u^{\varphi })^2\\
&-r \{2 M (u^t-a u^{\varphi })^2+r [(a^2+r^2) (u^{\varphi })^2-(u^t)^2]\}=r^2,\nn\\
\label{w4}&Q^2 (u^t-a u^{\varphi })^2\\
&+r [r^3 (u^{\varphi })^2-M (u^t-a u^{\varphi })^2]=-e r^3 (F_{rt}
u^t+F_{r\varphi } u^{\varphi }),\nn
\end{align}
where $F_{rt}$ and $F_{r\varphi }$ are given by~\eqref{ef2} taking $f=Mr-Q^2/2$ ($f_1=M$ and $f_2=-Q^2/2$). Numerical solutions to Eqs.~\eqref{w3}, \eqref{w4}, and~\eqref{f5}, with $\omega_{r}^2$ being given by~\eqref{p8b}, have been obtained and plots of $r_{\text{isco}}$ and $\omega_{\text{isco}}$ ($=u^{\varphi}/u^t$) in terms of $a$ are depicted in Fig.~\ref{Fig1} (the case of prograde circular orbits, $u^{\varphi}>0$, of a charged particle) and in Fig.~\ref{Fig2} (the case of retrograde circular orbits, $u^{\varphi}<0$, of a charged particle) for different values of $\mathcal{B}$. For all the plots, but the plots of figures~\ref{Fig3} and~\ref{Fig10}, we have fixed
\begin{align}
&M=1,\quad  0\leq a\leq 0.02,\quad 0\leq\mathcal{B}\leq 1,\nn\\
\label{w4b}&Q=+1/10,\quad \mathcal{Q}= +1/500\qquad (\text{black plots}),\\
&Q=-1/10,\quad \mathcal{Q}= -1/500\qquad (\text{grey plots}),\nn
\end{align}
in the system of units where $c=G=1/(4\pi\epsilon_0)=1$. The value $\mathcal{B}=1$ is considered as an upper limit. As we stated earlier, the expressions~\eqref{ef2} are linear in $a$ and are not valid for the whole range of $a$. For the case of the Kerr black hole immersed in a magnetic field some authors have, however, considered the whole range of $a$. We do not expect the results to remain valid in the limit $a\to 1$, particularly the conclusions drawn concerning the Aschenbach effect.

On comparing Figs.~\ref{Fig1} and~\ref{Fig2} one sees that for prograde orbits $r_{\text{isco}}$ decreases and $\omega_{\text{isco}}$ increases with $a$ if $\mathcal{B}$ is held constant. For retrograde orbits $r_{\text{isco}}$ increases and $|\omega_{\text{isco}}|$ decreases with $a$ if $\mathcal{B}$ is held constant. Now, if $a$ is held constant, $r_{\text{isco}}$ decreases for prograde orbits and it increases for retrograde orbits as $\mathcal{B}$ increases. All that is independent of the sign of $\mathcal{Q}$~\eqref{w4b} (with $Q\mathcal{Q}$ being positive).

For $a$ held constant, one sees from Fig.~\ref{Fig1} (corresponding to prograde orbits) that the gap between the two plots corresponding to $\mathcal{Q}=+1/500$ and $\mathcal{Q}=-1/500$ narrows as $\mathcal{B}$ increases. This same gap enlarges for retrograde orbits as $\mathcal{B}$ increases. To make this last statement more obvious we provide the values of $r_{\text{isco}}$ and $\omega_{\text{isco}}$ at $a=0$ corresponding to $\mathcal{Q}=+1/500$ and $\mathcal{Q}=-1/500$, respectively, with more decimal places than those shown in Fig.~\ref{Fig2}: for $\mathcal{B}=0$, $r_{\text{isco}}=5.97647,\ 5.99346$ and $\omega_{\text{isco}}=-0.068338,\ -0.068144$, for $\mathcal{B}=0.001$, $r_{\text{isco}}=6.26805,\ 6.28805$ and $\omega_{\text{isco}}=-0.063989,\ -0.063777$, for $\mathcal{B}=0.003$, $r_{\text{isco}}=7.74603,\ 7.80349$ and $\omega_{\text{isco}}=-0.047502,\ -0.047066$.

\emph{Radially stable prograde circular orbits exit for the whole range of $\mathcal{B}$~\eqref{w4b} and it seems that there is some critical value $\mathcal{B}_c$ beyond which no radially stable retrograde circular orbit exists (in the vicinity of $r_{\text{isco}}$ of the prograde circular orbits); however, as we shall see in Sec.~\ref{secqpos}, vertically stable circular orbits do exist. For $Q=1/10$ and $\mathcal{Q}=1/500$~\eqref{w4b} we find $\mathcal{B}_c=0.003204$ and for $Q=-1/10$ and $\mathcal{Q}=-1/500$ we find $\mathcal{B}_c=0.003189$. That's why we have chosen $\mathcal{B}\leq 0.003$ in Fig.~\ref{Fig2}.} This statement, which we will improve in Sec.~\ref{seccon}, remains valid in the vicinity of $r_{\text{isco}}$.

By the constraint that $|\mathcal{Q}|\ll 1$~\eqref{w4b} we assume that accretion disks do carry electric charges of both signs the average values of which remain relatively small. For elementary particles, however, $|\mathcal{Q}|$ is too large~\cite{Kerr1}. Had we chosen $\mathcal{Q}=+500$ we would have obtained the plots of Fig.~\ref{Fig3}. For $\mathcal{Q}<0$ ($eQ<0$) it is straightforward to prove analytically the existence of circular orbits. For $\mathcal{Q}>0$ ($eQ>0$) it is also possible, at least for small $a$, to set the conditions for the existence of circular orbits, however, since these conditions are necessary (not sufficient) it is not worth writing them. The plots of Fig.~\ref{Fig3} show that circular orbits exist for $0\leq \mathcal{B}\leq 1$ and $\mathcal{Q}>0$. Contrary to the case $|\mathcal{Q}|\ll 1$~\eqref{w4b}, for large and positive $\mathcal{Q}$ it seems that there is a critical value $\bar{\mathcal{B}}_c$ beyond which no prograde circular orbit exists. For $\mathcal{Q}=+500$, we find $\bar{\mathcal{B}}_c=0.110000$.

If the test particle is uncharged ($e=0$), the magnetic field will have no effect on the motion of the particle and the problem reduces to that of of an uncharged particle in the geometry of the Kerr--Newman black hole. The determination of ($r_{\text{isco}},\,\omega_{\text{isco}}$) can be carried analytically resulting in
\begin{align}
\label{w5a}&\omega_{\text{isco}}=\frac{\sqrt{M r_{\text{isco}}-Q^2}}{a\sqrt{M r_{\text{isco}}-Q^2}\pm r_{\text{isco}}^2},\\
&M r_{\text{isco}}^3-6 M^2 r_{\text{isco}}^2+3 M (3 Q^2-a^2) r_{\text{isco}}+4 Q^2 (a^2-Q^2)\nn\\
\label{w5}&\pm 8 a (M r_{\text{isco}}-Q^2)^{3/2}=0,
\end{align}
where the ``$+$" sign corresponds to prograde circular orbits and the ``$-$" sign corresponds to retrograde circular orbits and where $r_{\text{isco}}$ is the smallest positive root of the algebraic equation~\eqref{w5} that is larger than the event horizon. This constitutes an analytic solution to Eqs.~\eqref{w3}, \eqref{w4}, and~\eqref{f5}, with $e=0$ and $\omega_{r}^2$ being given by~\eqref{p8b}. Equation~\eqref{w5} is known in the scientific literature~\cite{Chandra,cr1}\footnote{The sign ``$\mp$" in Eq.~(81) of Ref.~\cite{cr1} should read ``$\pm$".}.

\subsubsection{Epicyclic frequencies of charged particles in the Kerr--Newman black hole immersed in a magnetic field\label{secqpos}}
In this section we focus on the effects of the magnetic filed on the stability of the circular motion. We fix the values of $M=1$, $Q=1/10$, $a=1/100$ and $\mathcal{Q}= 1/500$ as in~\eqref{w4b}, we solve numerically Eqs.~\eqref{w3} and~\eqref{w4}, and evaluate the epicyclic frequencies $\omega_{r}^2$~\eqref{p8b} and $\omega_{\theta}^2$~\eqref{p5b}.

\begin{figure*}
	\centering
	\includegraphics[width=0.33\textwidth]{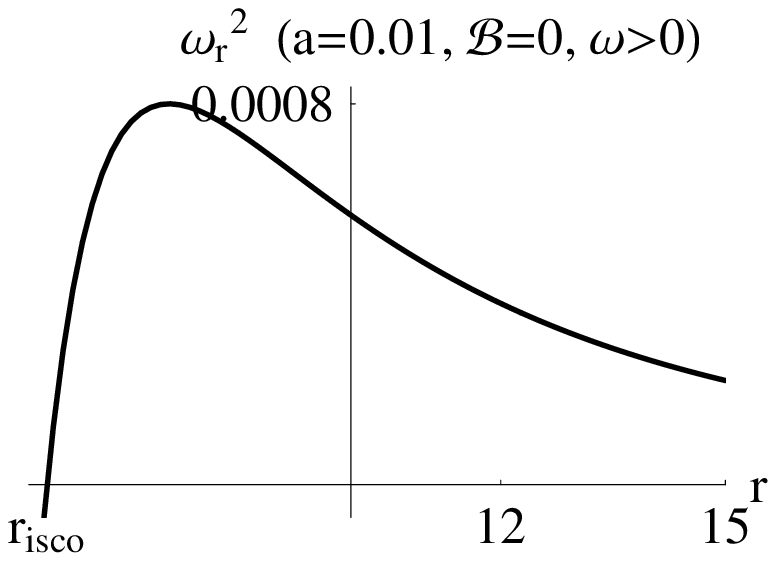} \includegraphics[width=0.33\textwidth]{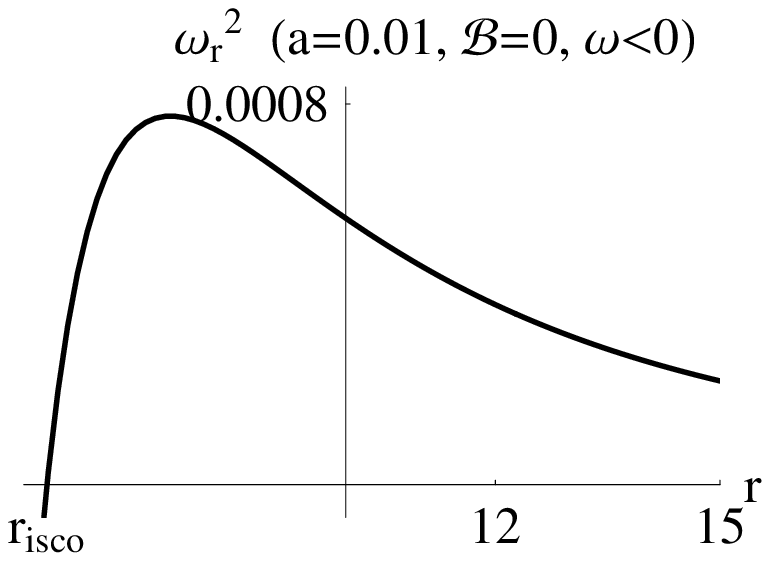} \\
	\includegraphics[width=0.33\textwidth]{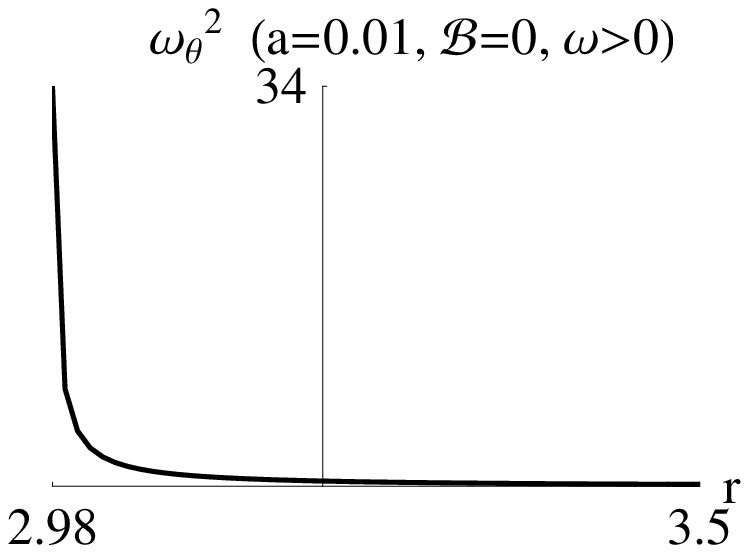} \includegraphics[width=0.33\textwidth]{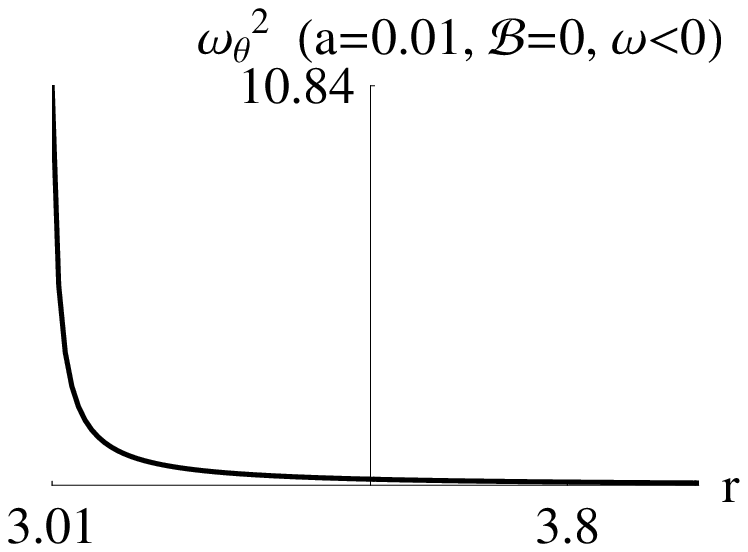} \\
	\caption{\footnotesize{Plots of the epicyclic frequencies $\omega_r^2$ and $\omega_{\theta}^2$ for $M=1$, $Q=1/10$, $a=1/100$, $\mathcal{Q}= 1/500$ and $\mathcal{B}=0$. This is the Kerr--Newman case. For the prograde orbits ($\omega>0$) and retrograde orbits ($\omega<0$) we have $r_{\text{isco}}=5.943754$ and $r_{\text{isco}}=6.009110$, respectively. This proves the existence of vertically stable, but radially unstable, circular motion in the vicinity of $r_{\text{isco}}$ ($r_h<r<r_{\text{isco}}$ where $r_h=1.994937$ is the event horizon).}}\label{Fig4}
\end{figure*}

\begin{figure*}
	\centering
	\includegraphics[width=0.33\textwidth]{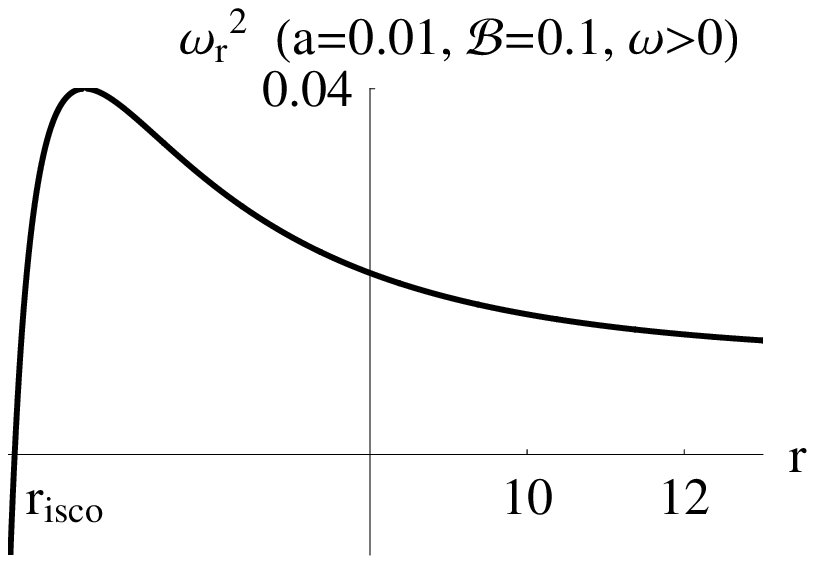} \includegraphics[width=0.33\textwidth]{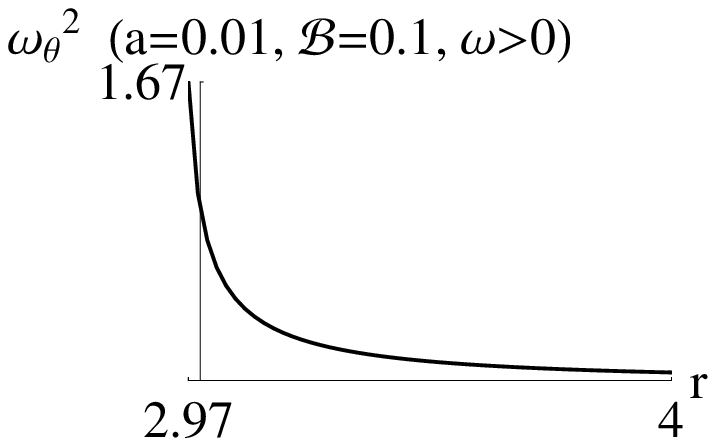} \includegraphics[width=0.33\textwidth]{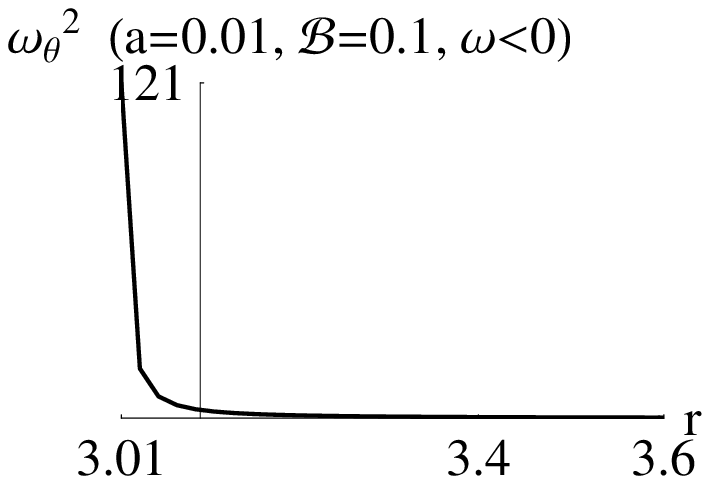} \\
	\caption{\footnotesize{Plots of the epicyclic frequencies $\omega_r^2$ and $\omega_{\theta}^2$ for $M=1$, $Q=1/10$, $a=1/100$, $\mathcal{Q}= 1/500$ and $\mathcal{B}=0.1$. For the prograde orbits ($\omega>0$) we have $r_{\text{isco}}=3.478461$ and this again proves the existence of vertically stable, but radially unstable, prograde circular motion in the vicinity of $r_{\text{isco}}$ ($r_h<r<r_{\text{isco}}$ where $r_h=1.994937$ is the event horizon). As we have stated in the caption of Fig.~\ref{Fig2} and in the second paragraph preceding the paragraph containing Eq.~\eqref{w5a}, there are no radially stable retrograde circular orbits ($\omega_r^2<0$) in the vicinity of $r_{\text{isco}}$ of the prograde circular orbits for $\mathcal{B}=0.1$.}}\label{Fig5}
\end{figure*}

\begin{figure*}
	\centering
	\includegraphics[width=0.33\textwidth]{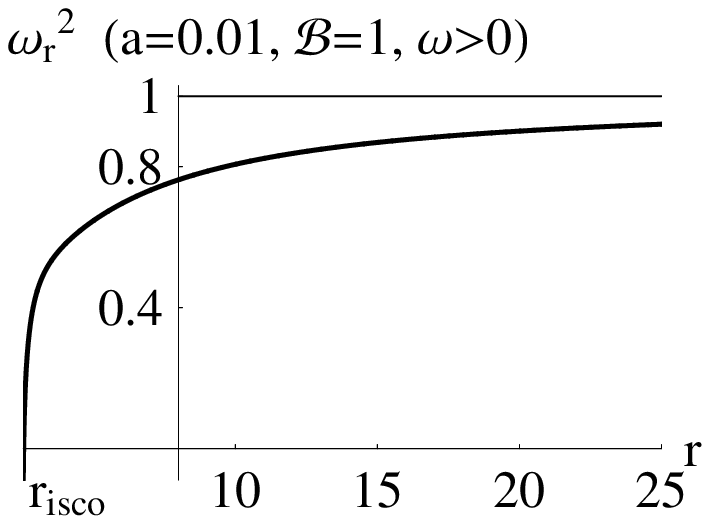} \includegraphics[width=0.33\textwidth]{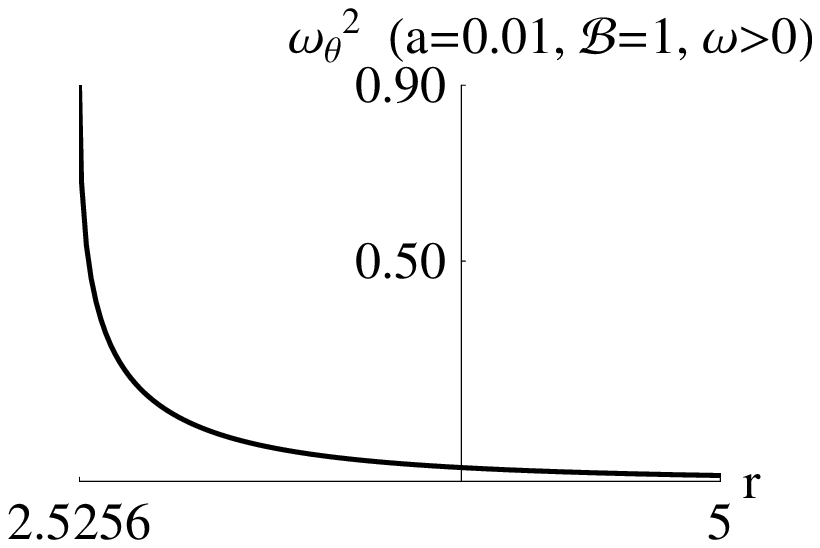} \includegraphics[width=0.33\textwidth]{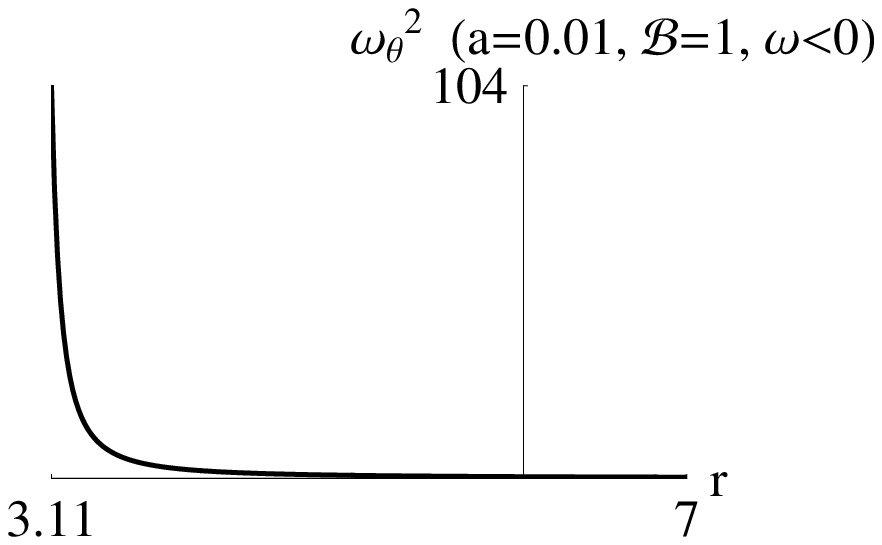} \\
	\caption{\footnotesize{Plots of the epicyclic frequencies $\omega_r^2$ and $\omega_{\theta}^2$ for $M=1$, $Q=1/10$, $a=1/100$, $\mathcal{Q}= 1/500$ and $\mathcal{B}=1$. For the prograde orbits ($\omega>0$) we have $r_{\text{isco}}=2.548148$. Here again we observe the existence of vertically stable, but radially unstable, prograde circular motion for $r_h<r<r_{\text{isco}}$ where $r_h=1.994937$ is the event horizon, but the region in which such orbits exit is narrower than it is in the cases of figures~\ref{Fig4} and~\ref{Fig5}. As we have stated in the caption of Fig.~\ref{Fig2} and in the second paragraph preceding the paragraph containing Eq.~\eqref{w5a}, there are no radially stable retrograde circular orbits ($\omega_r^2<0$) in the vicinity of $r_{\text{isco}}$ of the prograde circular orbits for $\mathcal{B}=1$. For prograde orbits, $\omega_r^2$ no longer approaches 0 in the limit $r\to\infty$, as is the case in figures~\ref{Fig4} and~\ref{Fig5}, rather it approaches 1. Moreover, the region, where vertically stable retrograde circular orbits exit, has been pushed entirely beyond $r_{\text{isco}}=2.548148$, starting from 3.11, and this was not the case in figures~\ref{Fig4} and~\ref{Fig5}. The leftmost and middle plots of this figure correspond to the branch $0<u^{\varphi}< 0.5881$ of the left plot of Fig.\ref{Fig7} and the rightmost plot of this figure corresponds to the branch $u^{\varphi}< 0$ of the left plot of Fig.\ref{Fig7}.}}\label{Fig6}
\end{figure*}

The first thing we prove, as shown in figures~\ref{Fig4} to~\ref{Fig6}, is the existence of a vertically stable, but radially unstable, prograde circular motion in the vicinity of $r_{\text{isco}}$ (for $r_h<r<r_{\text{isco}}$ where $r_h=1.994937$ is the event horizon). This means that there is some radius $r_c<r_{\text{isco}}$, closer to $r_{\text{isco}}$, where a prograde circular motion is still possible and $r_{\text{isco}}-r_c$ decreases with $\mathcal{B}$. As $\mathcal{B}$ increases the radially stable retrograde circular motion disappears but the vertically stable motion continues to exist in the vicinity of $r_{\text{isco}}$, as shown in the rightmost plots of figures~\ref{Fig5} and~\ref{Fig6}. Here $r_{\text{isco}}$ is the radius of the isco of the prograde circular orbits (that of retrograde circular orbits no ``longer exists").

If the dimensionless parameter $\mathcal{B}$~\eqref{w2} is relatively small, $\omega_{r}^2$ and $\omega_{\theta}^2$ approach 0 in the limit $r\to\infty$ as shown in figures~\ref{Fig4} and~\ref{Fig5}. For relatively high ratio $e$ or large black holes ($M$ large), $\mathcal{B}$ approaches (its upper limit) 1~\eqref{w4b} and in this case $\omega_{r}^2$ approaches 1, instead of 0, as shown in the leftmost plot of Fig.~\ref{Fig6}. Here again we see that there must exist some critical value of $\mathcal{B}$ beyond which $\omega_{r}^2$ no longer approaches 0 as $r\to\infty$.

This is not the whole story! For large $\mathcal{B}$~\eqref{w4b}, in the near vicinity of the horizon, new effects take place and the notion of isco becomes ambiguous. For $M=1$, $Q=1/10$, $a=1/100$, $\mathcal{Q}= 1/500$ and $\mathcal{B}=1$, the left plot of Fig.~\ref{Fig7} shows the behavior of the multivalued function $u^{\varphi}(r)$ in the vicinity of $r_{\text{isco}}=2.548148$. This plot is nothing but a representation of the plots depicted in Fig.~\ref{Fig6}. The circular motion corresponding to $0<u^{\varphi}< 0.5881$ is stable with epicyclic frequencies $\omega_r^2$ and $\omega_{\theta}^2$ depicted in the leftmost and middle plots of Fig.~\ref{Fig6}. The branch corresponding to $u^{\varphi}< 0$ is only vertically stable with $\omega_{\theta}^2$ depicted in the rightmost plot of Fig.~\ref{Fig6}. The branch corresponding to $u^{\varphi}\geq 0.5881$ is only vertically stable.

The new effects take place in the region adjacent to the event horizon $r_{h}=1.994937$, as shown in the right plot of Fig.~\ref{Fig7}, which is a plot of $u^{\varphi}(r)$ for the same values of the parameters and $r$ runs in the too-narrow region, $r_h<1.994987<r<1.995024$, that could not be fitted in the left plot. First notice the presence of a point $r=1.995015$ where $u^{\varphi}=0$: The charged particle remains at rest on the circle of radius $r=1.995015$, which is radially stable with $\omega_r^2=74997.8$ and vertically unstable with $\omega_{\theta}^2=-0.21577$.

The right plot of Fig.~\ref{Fig7} could be divided into three branches: (a) $u^{\varphi}\geq 0.3818$, (b) $0\leq u^{\varphi}< 0.3818$, and (c) $u^{\varphi}<0$. The branch (a) is stable where $\omega_{r}^2$ is depicted in the leftmost plot of Fig.~\ref{Fig8} and $\omega_{\theta}^2$ is depicted in the upper-left plot of Fig.~\ref{Fig9}. The branch (b) is only radially stable where $\omega_{r}^2$ is depicted in the middle plot of Fig.~\ref{Fig8} and $\omega_{\theta}^2$ is depicted in the upper-right plot of Fig.~\ref{Fig9}. The branch (c) is stable for $r\leq 1.995011$ where $\omega_{r}^2$ is depicted in the rightmost plot of Fig.~\ref{Fig8} and $\omega_{\theta}^2$ is depicted in the two lower plots of Fig.~\ref{Fig9}. We have thus proven the existence of other stable \emph{prograde and retrograde} circular orbits in the region enclosed by the circles $r=r_h$ and $r=r_{\text{isco}}$. This shows that there is no isco or that the notion of isco becomes ambiguous for large $\mathcal{B}$. For retrograde circular orbits, this ambiguity may be solved by defining a new $\theta$-isco corresponding to $\omega_{\theta}^2=0$ yielding $r_{\text{isco}}=1.995011$, as shown in the lower-right plot of Fig.~\ref{Fig9} (this is just a zoomed in depiction of the lower-left plot of Fig.~\ref{Fig9}).

\section{Conclusion\label{seccon}}
We have presented a general procedure to decouple the perturbed equations of a circular motion of a charged particle in the geometry of a stationary solution (that could be a rotating black hole, wormhole, etc.). The epicyclic frequencies have been given in a metric-independent and electromagnetic-field-independent forms.

\begin{figure*}
	\centering
	\includegraphics[width=0.33\textwidth]{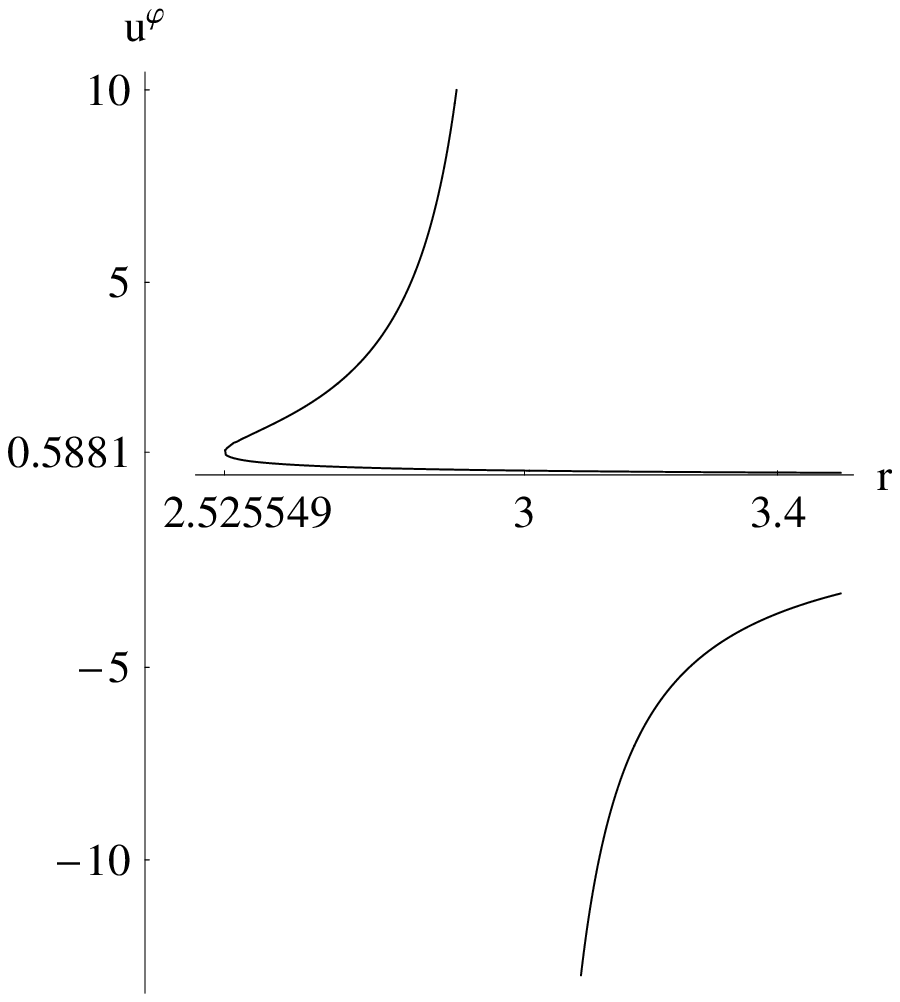} \includegraphics[width=0.33\textwidth]{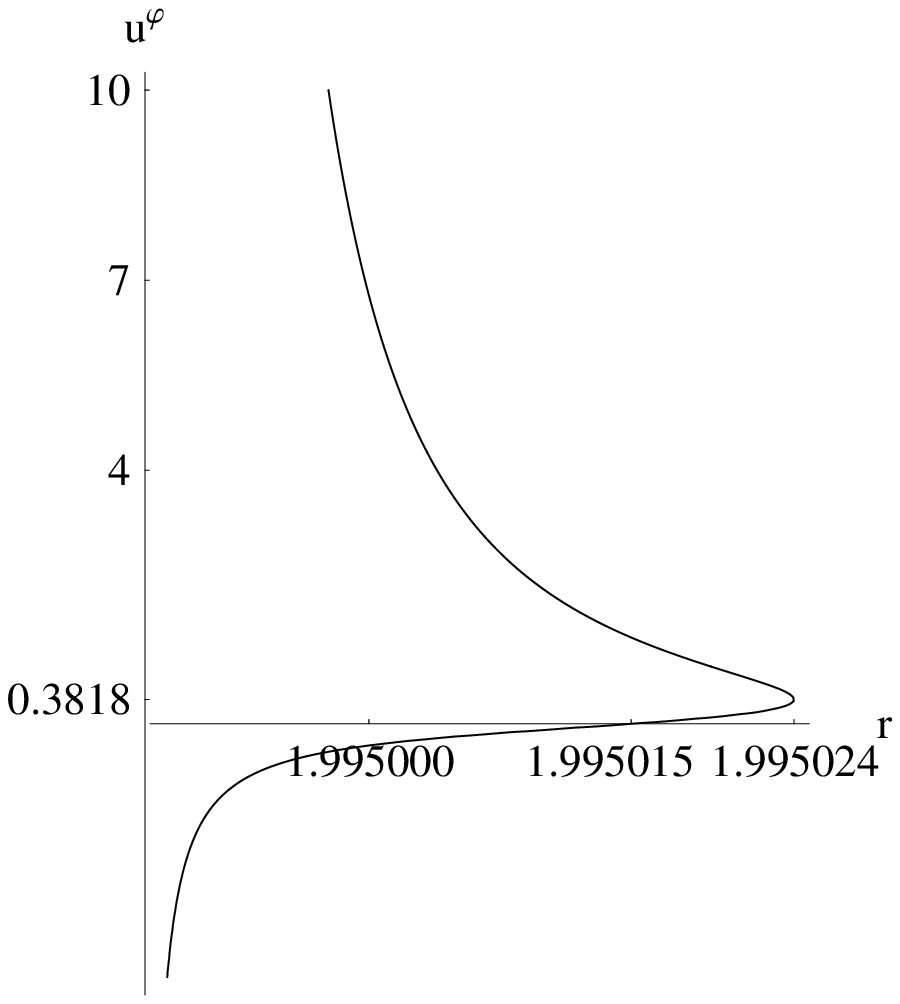} \\
	\caption{\footnotesize{Plots of the multivalued function $u^{\varphi}$ versus $r$ for $M=1$, $Q=1/10$, $a=1/100$, $\mathcal{Q}= 1/500$ and $\mathcal{B}=1$. \textit{Left Plot}: It shows the behavior of $u^{\varphi}(r)$ in the vicinity of $r_{\text{isco}}=2.548148$. The circular motion corresponding to $0<u^{\varphi}< 0.5881$ is stable with epicyclic frequencies $\omega_r^2$ and $\omega_{\theta}^2$ depicted in the leftmost and middle plots of Fig.~\ref{Fig6}. The branch corresponding to $u^{\varphi}< 0$ is only vertically stable with $\omega_{\theta}^2$ depicted in the rightmost plot of Fig.~\ref{Fig6}. The branch corresponding to $u^{\varphi}\geq 0.5881$ is only vertically stable. \textit{Right Plot}: This is a zoomed in depiction of the left plot. It shows the behavior of $u^{\varphi}(r)$ in the near vicinity of the event horizon $r_{h}=1.994937$ (the $u^{\varphi}$ axis in this plot passes through the event horizon of the Reissner-Nordstr\"om, $M+\sqrt{M^2-Q^2}=1.994987$, which is slightly larger than $r_h=1.994937$). This plot could be divided into three branches: (a) $u^{\varphi}\geq 0.3818$, (b) $0\leq u^{\varphi}< 0.3818$, and (c) $u^{\varphi}<0$. The branch (a) is stable where $\omega_{r}^2$ is depicted in the leftmost plot of Fig.~\ref{Fig8} and $\omega_{\theta}^2$ is depicted in the upper-left plot of Fig.~\ref{Fig9}. The branch (b) is only radially stable where $\omega_{r}^2$ is depicted in the middle plot of Fig.~\ref{Fig8} and $\omega_{\theta}^2$ is depicted in the upper-right plot of Fig.~\ref{Fig9}. The branch (c) is stable for $r\leq 1.995011$ where $\omega_{r}^2$ is depicted in the rightmost plot of Fig.~\ref{Fig8} and $\omega_{\theta}^2$ is depicted in the two lower plots of Fig.~\ref{Fig9}.}}\label{Fig7}
\end{figure*}

\begin{figure*}
	\centering
	\includegraphics[width=0.33\textwidth]{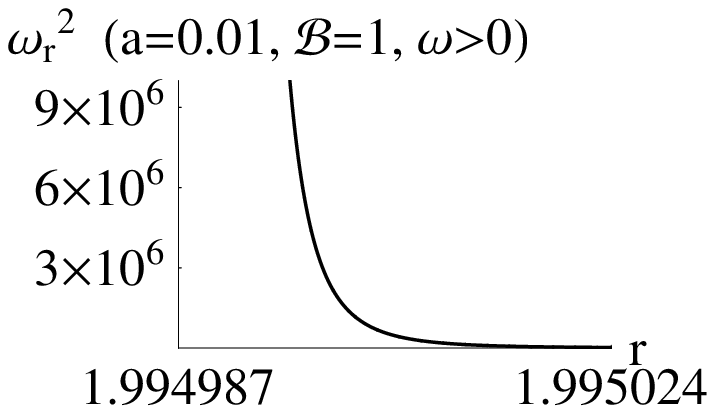} \includegraphics[width=0.33\textwidth]{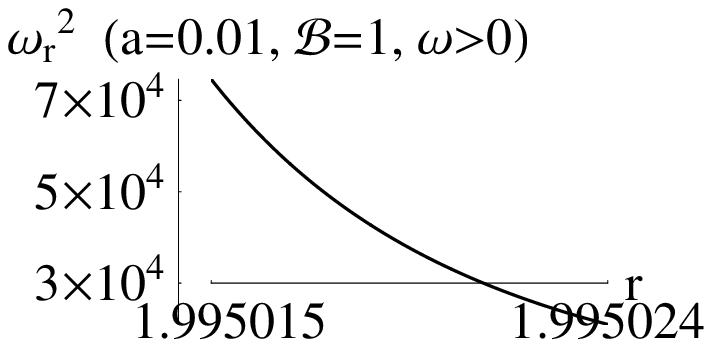} \includegraphics[width=0.33\textwidth]{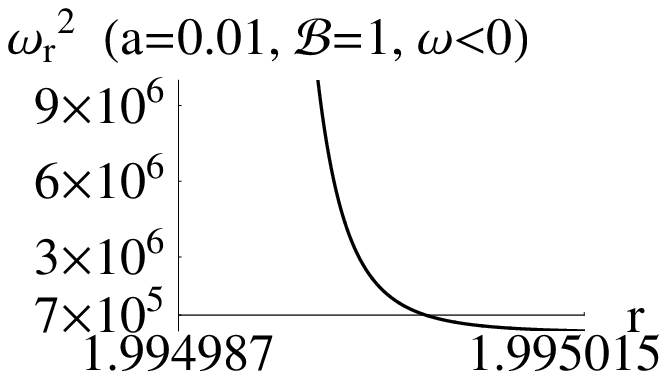} \\
	\caption{\footnotesize{Plots of the epicyclic frequency $\omega_r^2$ for $M=1$, $Q=1/10$, $a=1/100$, $\mathcal{Q}= 1/500$ and $\mathcal{B}=0.1$. From left to right, these plots correspond to the branches (a) $u^{\varphi}\geq 0.3818$, (b) $0\leq u^{\varphi}< 0.3818$, and (c) $u^{\varphi}<0$ of the right plot of Fig.~\ref{Fig7}, respectively.}}\label{Fig8}
\end{figure*}

\begin{figure*}
	\centering
	\includegraphics[width=0.33\textwidth]{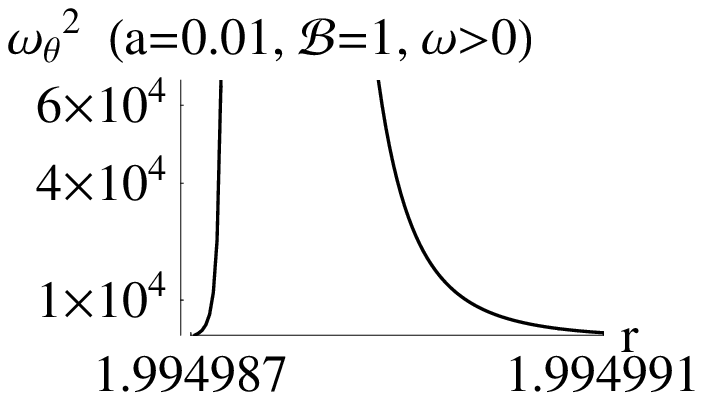} \includegraphics[width=0.33\textwidth]{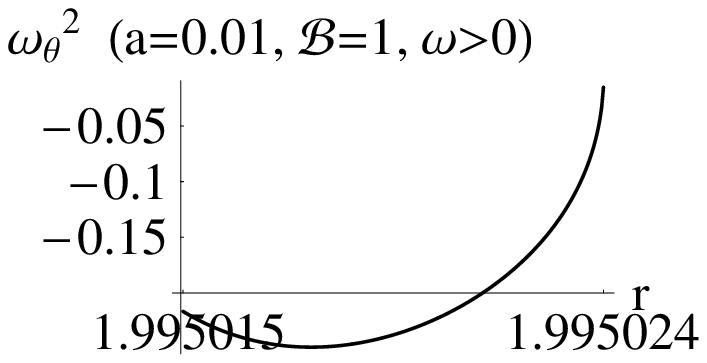} \\ \includegraphics[width=0.33\textwidth]{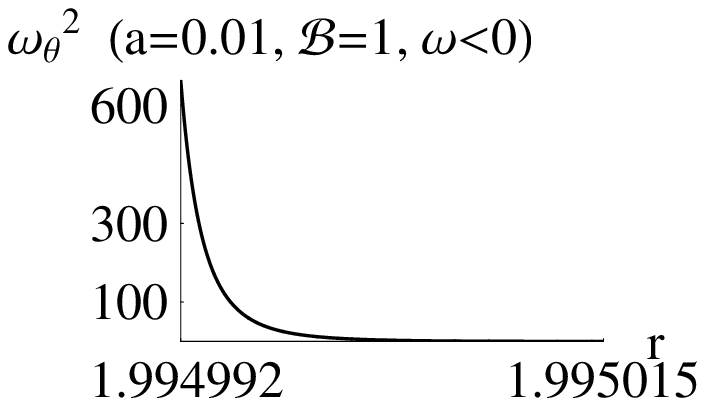} \includegraphics[width=0.33\textwidth]{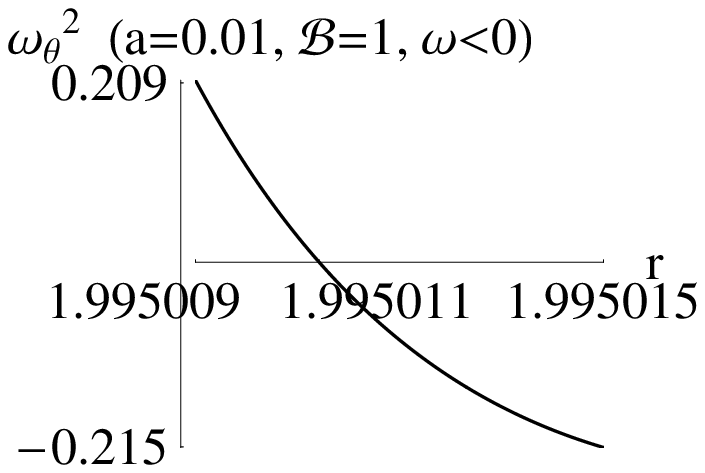} \\
	\caption{\footnotesize{Plots of the epicyclic frequency $\omega_{\theta}^2$ for $M=1$, $Q=1/10$, $a=1/100$, $\mathcal{Q}= 1/500$ and $\mathcal{B}=0.1$. The upper-left plot corresponds to the branch (a) $u^{\varphi}\geq 0.3818$ of the right plot of Fig.~\ref{Fig7} (there is no singularity in the value of $\omega_{\theta}^2$ for $1.994987<r<1.994991$, for the sake of clarity we have avoided to represent the range of $\omega_{\theta}^2$), the upper-right plot corresponds to the branch (b) $0\leq u^{\varphi}< 0.3818$ of the right plot of Fig.~\ref{Fig7}, and the the lower plots correspond to the branch (c) $u^{\varphi}<0$ of the right plot of Fig.~\ref{Fig7}. The lower-right plot of this figure is just a zoomed in depiction of the lower-left plot.}}\label{Fig9}
\end{figure*}

Applications to the Kerr--Newman black hole immersed in a magnetic field have been considered. We have shown that in the presence of strong parameter $\mathcal{B}$ there may be two types of iscos for circular orbits, the usual $r$-isco corresponding to $\omega_{r}^2=0$ and the $\theta$-isco corresponding to $\omega_{\theta}^2=0$, as there may be no isco at all. Investigation of the plots in Fig.~\ref{Fig6} reveals the existence of a region bounded above by the isco and adjacent to it where circular motion, however only vertically stable, is still possible. This effect is already known for the Kerr black hole immersed in a magnetic field. By the discovery we have made in the right plot of Fig.~\ref{Fig7}, we have extended the analysis of the epicyclic frequencies to the region adjacent to the event horizon (bounded above by the usual isco). 

From all the cases we have investigated we may draw the following conclusions: It is possible that all the circular orbits are grouped into circular bands around the source with an isco in each band with the ``i" refereing to ``inner" instead of ``innermost". We have only considered the case $\mathcal{B}=1$, which is a relatively high value, some authors considered much higher values. It is interesting to investigate all possible effects of a strong parameter $\mathcal{B}$.

One can now improve the statement made in the second paragraph preceding the paragraph containing Eq.~\eqref{w5a}. \emph{Radially and vertically stable prograde and retrograde circular orbits exit for the whole range of $\mathcal{B}$~\eqref{w4b} we have chosen; If they are not available in the region bounded below by the usually defined isco one should look for them in the region bounded above by the isco and below by the event horizon.} We have checked that some of these circular orbits exist in the region bounded above by the event horizon.

For the Kerr black hole we had tried hard to obtain the same result, that is, the same depiction as the right plot of Fig.~\ref{Fig7} but we could not observe anything similar even for $\mathcal{B}=10$ and $\mathcal{B}=100$; All we obtained were depictions similar to the left plot of Fig.~\ref{Fig7} that are more displaced towards the event horizon as $\mathcal{B}$ is increased.

\begin{figure*}
	\centering
	\includegraphics[width=0.33\textwidth]{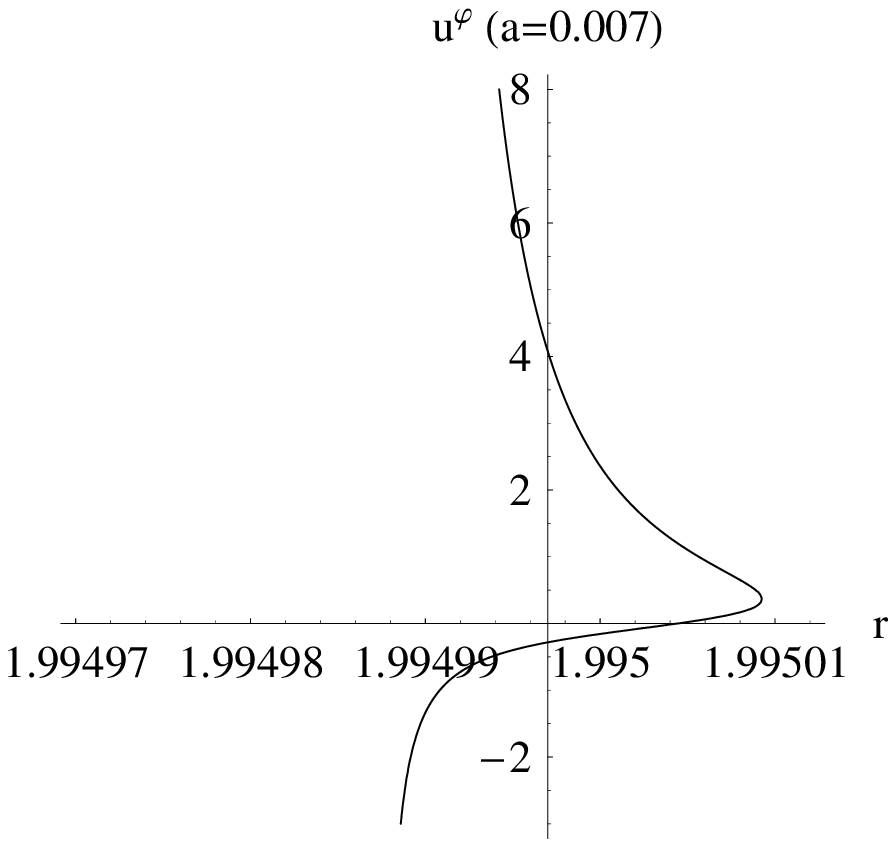} \includegraphics[width=0.33\textwidth]{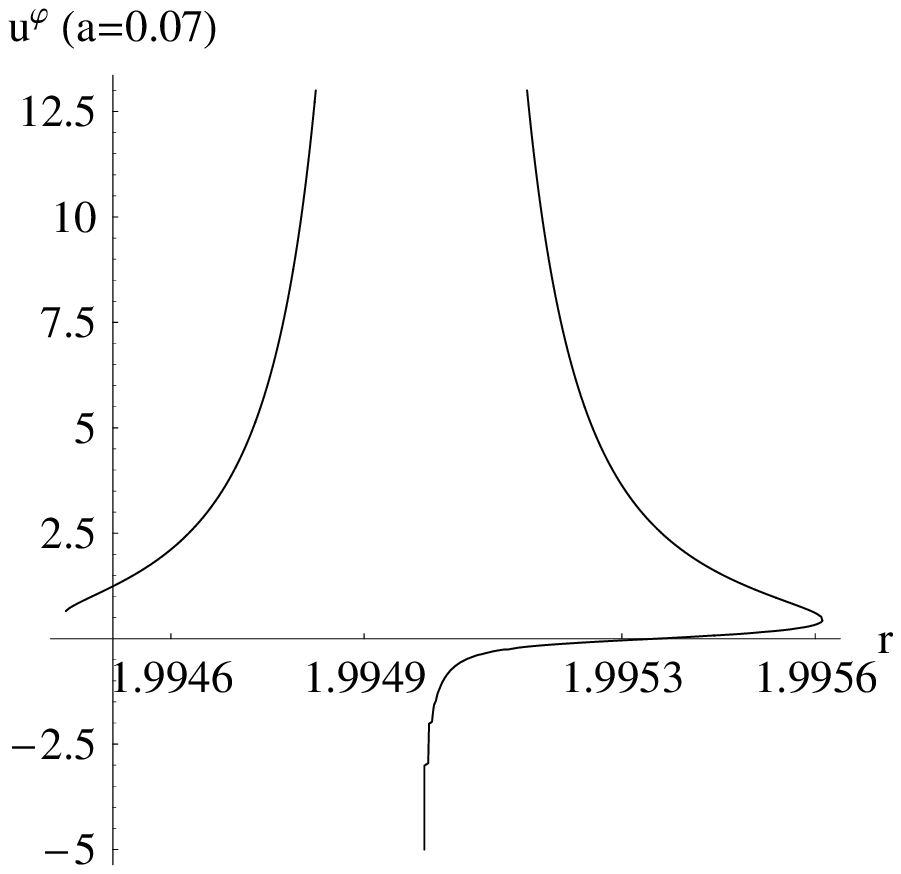} \\ \caption{\footnotesize{Plots of the multivalued function $u^{\varphi}$ versus $r$ for $M=1$, $Q=1/10$, $\mathcal{Q}= 1/500$ and $\mathcal{B}=1$. Here we are depicting $u^{\varphi}(r)$ in the near vicinity of the event horizon; the other branch adjacent to isco still exists and is very similar to the left plot of Fig.~\ref{Fig7}. \textit{Left Plot}: $a=7/1000$. It shows the behavior of $u^{\varphi}(r)$ in the near vicinity of the event horizon $r_{h}=1.99496$. \textit{Right Plot}: $a=7/100$. It shows the behavior of $u^{\varphi}(r)$ in the near vicinity of the event horizon $r_{h}=1.9925$. Notice the presence of a second branch almost symmetric, but truncated near the event horizon, to the already existing one in the left plot of this figure and in the right plot of Fig.~\ref{Fig7}.}}\label{Fig10}
\end{figure*}

To show how rotation affects the behavior of $u^{\varphi}(r)$ in the vicinity of the event horizon,  we generated in Fig.~\ref{Fig10} a similar plot to that of Fig.~\ref{Fig7}. For $a=0.007$ and $a=0.07$ (which is still a relatively small value) we obtained the cases depicted in Fig.~\ref{Fig10} where only the branch of $u^{\varphi}(r)$ that is adjacent to the event horizon is represented (the other branch adjacent to isco is very similar to the left plot of Fig.~\ref{Fig7}). As $a$ increases, the graph of $u^{\varphi}(r)$ develops another branch almost symmetric, but truncated near the event horizon, to the already existing one.

We have also shown that a relatively strong parameter $\mathcal{B}$ makes black holes better particle accelerators by stabilizing both \emph{prograde and retrograde} circular orbits in the near vicinity of the event horizon. Such accelerators are able to project particles of the same charge sign in opposite directions and collide them after some turns around the black hole. Static and nearly static (nonrelativistic) observers in the near vicinity of the horizon are possible.


%
\section*{Appendix: Nonvanishing expressions of $F^{\mu\nu}$\label{secaa}}
\renewcommand{\theequation}{A.\arabic{equation}}
\setcounter{equation}{0}
\begin{align}
&F^{t r}=\frac{Q}{r^2}-\frac{B a (2 f_2+f_1 r) (1+3 \cos  2 \theta )}{2 r^3},\nn\\
&F^{t \theta }=-\frac{B a f_2 \sin  2 \theta }{r^4},\nn\\
\label{A1}&F^{r \varphi }=\frac{B (r^2-2 f)}{r^3}-\frac{Q a}{r^4},\\
&F^{\theta \varphi }=\frac{B (2 f_2+r^2) \cot \theta }{r^4}+\frac{2 Q a \cot \theta }{r^5}.\nn
\end{align}



\end{document}